\documentclass[article,11pt]{article}

\pdfoutput=1

\usepackage{jheppub}
\usepackage{caption}
\usepackage{subcaption}
\usepackage{amssymb}
\usepackage{amsbsy}
\usepackage{amsfonts}
\usepackage{amsmath}
\usepackage{epsfig}
\usepackage{cancel}
\usepackage{slashed}
\usepackage{array}

\newcolumntype{L}[1]{>{\raggedright\let\newline\\\arraybackslash\hspace{0pt}}m{#1}}
\newcolumntype{C}[1]{>{\centering\let\newline\\\arraybackslash\hspace{0pt}}m{#1}}
\newcolumntype{R}[1]{>{\raggedleft\let\newline\\\arraybackslash\hspace{0pt}}m{#1}}

\title{LHC constraints on Mini-Split anomaly and gauge mediation and prospects for LHC 14 and a future 100 TeV $pp$ collider}

\author[a]{Hugues Beauchesne}
\author[a]{Kevin Earl}
\author[a]{Thomas~Gr\'egoire}

\emailAdd{HuguesBeauchesne@cmail.carleton.ca}
\emailAdd{KevinEarl@cmail.carleton.ca}
\emailAdd{gregoire@physics.carleton.ca}

\affiliation[a]{Ottawa-Carleton Institute for Physics, Department of Physics, Carleton University 1125 Colonel By Drive, Ottawa, K1S 5B6 Canada}
\abstract{Stringent experimental constraints have raised the lower limit on the masses of squarks to TeV levels, while compatibility with the mass of the Higgs boson provides an upper limit. This two-sided bound has lead to the emergence of Mini-Split theories where gauginos are not far removed from the electroweak scale while scalars are somewhat heavier. This small hierarchy  modifies the spectrum of standard anomaly and gauge mediation, leading to Mini-Split deflected anomaly and gauge mediation models. In this paper, we study LHC constraints on these models and their prospects at LHC 14 and a 100 TeV collider. Current constraints on their parameter space come from ATLAS and CMS supersymmetry searches, the known mass of the Higgs boson, and the absence of a color-breaking vacuum. Prospects at LHC 14 and a 100 TeV collider are obtained from these same theoretical constraints in conjunction with background estimates. As would be expected from renormalization group effects, a slightly lighter third generation of squarks is assumed. Higgsinos have masses similar to those of the scalars and are at the origin of the deflection.}

\begin{document}

\maketitle

\section{Introduction}
Supersymmetric (SUSY) extensions of the standard model (SM) at the weak scale have the advantages of solving the hierarchy problem, providing a dark matter candidate, and leading to gauge coupling unification. Even though weak scale SUSY is by no mean ruled out, lack of results at the LHC forces one to reconsider whether the Higgs mass might be fine-tuned to a certain degree. This is the case in Split-SUSY models \cite{Wells:2003tf, ArkaniHamed:2004fb, Giudice:2004tc} where fermion superpartners can still be close to the electroweak scale while  the scalar superpartners are much heavier. These theories are no longer solutions to the hierarchy problem (which could be explained by an environmental selection principle for example), but maintain a dark matter candidate and can keep intact gauge coupling unification \cite{Wells:2003tf, ArkaniHamed:2004fb, Giudice:2004tc}. However, it was shown \cite{Arvanitaki:2012ps} that scalars heavier than $10^5$ TeV would make it difficult to reconcile Split-SUSY with the known mass of the Higgs boson \cite{Chatrchyan:2013mxa, Aad:2014aba}, therefore putting an upper limit on this splitting. These Split-SUSY theories with only a small gap are referred to as Mini-Split \cite{Arvanitaki:2012ps}. 

One of the main phenomenological characteristics of Mini-Split models is the presence of a small hierarchy between the gauginos and the scalars. The  conventional gaugino mass spectra associated to well-known mediation mechanisms like anomaly mediation \cite{Randall:1998uk, Giudice:1998xp} and gauge mediation \cite{Dine:1981gu, Nappi:1982hm, AlvarezGaume:1981wy, Dine:1993yw, Dine:1994vc, Dine:1995ag} are then modified, as the heavy superpartners deflect the gaugino masses from their standard renormalization group (RG) trajectory when they are integrated out. The resulting spectra are referred to as deflected anomaly mediation \cite{ Rattazzi:1999qg, Bagnaschi:2014rsa} or deflected gauge mediation. The precise  phenomenology of  Mini-Split models depends on the value of  $\mu$ which could either be at the electroweak scale or at the same scale as the scalars. In this work, we focus on the case of large $\mu$. The case of small $\mu$ was considered in \cite{Jung:2013zya} which provides future prospects for anomaly mediation in Mini-Split theories at a 100 TeV collider with light Higgsinos (which minimizes the amount of deflection) and applies these results to gauge and mirror mediation. Reference \cite{Jung:2013zya} also studied cases with a large $\mu$ (50 TeV) but still somewhat smaller than what is considered in most of the parameter space we consider. Dark matter predictions for such models are presented in \cite{Cesarini:2006jp, Yokozaki:2009fu}. Other variants of deflected mediation are studied in \cite{Setzer:2010bz, Okada:2002mv, Okada:2012nr, deBlas:2011cr, Ibe:2006de, Ibe:2011aa, Ibe:2012hu, Bhattacherjee:2012ed, Kahn:2013pfa}. 

The purpose of this paper is to constrain the parameter space of Mini-Split models with deflected anomaly and gauge mediation  using LHC data and to predict future exclusion and discovery prospects at LHC 14 and a future 100 TeV collider. Current constraints are extracted from ATLAS \cite{TheATLAScollaboration:2013zia, Aad:2014vma, Aad:2014nua, Aad:2014pda, Aad:2014lra} and CMS \cite{Khachatryan:2014mma, Chatrchyan:2014lfa, CMS:2013ija} SUSY searches (mainly gluino pair production), the known mass of the Higgs boson \cite{Chatrchyan:2013mxa, Aad:2014aba}, and the absence of a color-breaking vacuum \cite{Bagnaschi:2014rsa}. Future prospects for LHC 14 and a 100 TeV collider are obtained by using the same theoretical tools in conjunction with background estimates. In the cases studied here, the deflection comes mainly from the Higgsino sector \cite{Bagnaschi:2014rsa, ArkaniHamed:2012gw}, which is assumed to be around the scalar scale and the light neutralinos/charginos are almost pure gauginos. As one generally expects the third generation of squarks to be lighter because of renormalization group effects for example, this paper makes the simplifying assumption of a slightly lighter third generation. 

This paper is organized as follows. The necessary theoretical elements are presented first. This includes an explanation of how Mini-Split theories can arise in both anomaly and gauge mediation, as well as pole mass expressions and branching fractions. The procedure necessary to calculate the Higgs mass is also presented. The methodology used in obtaining both current limits and future prospects is then explained. This includes the LHC searches used to determine current limits.  Finally, we present current LHC constraints and prospects at LHC 14 and a future 100 TeV collider.

\section{Theory}

\subsection{Mini-Split models}\label{SectionMiniSplitModels}
In this section, we review how Mini-Split spectra can be realized in both anomaly and gauge mediation (see for example \cite{Bagnaschi:2014rsa}). Quite generally, sfermions masses can be generated via terms of the form
\begin{equation}\label{Squarkmassesgeneration}
   \int d^4 \theta \frac{X^{\dagger}X}{M_{\ast}^2}Q^{\dagger}Q,
\end{equation}
where $M_{\ast}$ is the mediation scale, $X=\theta^2 F_{X}$ is a SUSY breaking spurion, and $Q$ is a chiral superfield. This term is always allowed by symmetries, irrespective of the R-charge of $X$ or its gauge quantum numbers. On the other hand  gaugino masses are generated via terms of the form
\begin{equation}
   \int d^2 \theta \frac{X}{M_{\ast}}W_{i\alpha}W^\alpha_i,
\end{equation}
where $W_{i\alpha} (i = 1,2,3)$ are the gauge-strength superfields. Contrary to the sfermion masses of (\ref{Squarkmassesgeneration}),  here $X$ is required to be a singlet under all gauge and global charges in order for this term to be allowed. It is therefore easier to forbid, and in the models that we consider we assume that it is absent. There is however an unavoidable contribution to gaugino masses coming from anomaly mediation
\begin{equation}
   M_i=\frac{\beta_i}{g_i}m_{3/2}.
\end{equation}
The A-terms are also generated by anomaly mediation and are given by
\begin{equation}\label{eqAterm}
   A_y=-\frac{\beta_y}{y}m_{3/2},
\end{equation}
where $y$ is the corresponding Yukawa and $\beta_y$ is its beta function. A $B_\mu$ term can be generated by a term of the form
\begin{equation}
   \int d^4 \theta \frac{X^{\dagger}X}{M_{\ast}^2}H_u H_d.
\end{equation}
In Mini-Split scenarios, the $\mu$ term can either be large (at the scale of the scalars) or small (at the scale of the  gauginos) depending on how it is generated. In this work we concentrate on the case where it is large, which could be generated through the Giudice-Masiero mechanism \cite{Giudice:1988yz} where a term of the following form is introduced
\begin{equation}
   \int d^4\theta \Phi^{\dagger}\Phi\left[\hat{H}_{u,d}^{\dagger}\hat{H}_{u,d}+\left(c \hat{H}_u\hat{H}_d+\text{h.c}\right)\right].
\end{equation}
Here $c$ is an arbitrary dimensionless constant and $\Phi$ is the conformal compensator which gets a non-zero $F$-term once SUSY is broken: $\Phi = 1 - m_{3/2} \theta^2$ . Upon rescaling of the fields, this becomes
\begin{equation}
   \int d^4\theta \left[H_{u,d}^{\dagger}H_{u,d}+\left(c \frac{\Phi^{\dagger}}{\Phi}H_u H_d+\text{h.c}\right)\right]
\end{equation}
and  leads to a $\mu$ term, in addition to an additional contribution to $B_{\mu}$. These terms are of order $m_{3/2}$ and $m_{3/2}^2$ respectively. If gravity is the sole mediator of supersymmetry breaking, then $M_\ast$ is the Planck mass and this leads to the scalars and Higgsinos all having masses of roughly $m_{3/2}$ while the masses of the gauginos are a loop factor smaller, leading to a Mini-Split spectrum. The fact that the $\mu$ term is taken to be large will change the running of the gauge couplings compared to the more conventional split-spectrum with light Higgsinos. The prediction for $\alpha_s(M_Z)$  was found in \cite{ArkaniHamed:2012gw} to be smaller than with light Higgsinos, but still consistent with the measured value.

Gauge mediation can also lead to Mini-Split spectra. This can be done in a multitude of ways. We give an example taken from \cite{Arvanitaki:2012ps}. Assume a superpotential of the form
\begin{equation}
   W=M_R\left(\Phi_1\overline{\Phi}_1+\Phi_2\overline{\Phi}_2\right)+X\Phi_1\overline{\Phi}_2,
\end{equation}
where the $\Phi_i$ and the $\overline{\Phi}_i$ are messengers and $X=M+F\theta^2$ is a spurion that breaks SUSY and R-symmetry. This leads to gauginos masses of
\begin{equation}
\label{eq:gaugemediation}
   M_i=\frac{\alpha_i}{6\pi}\frac{M}{M_R}\frac{F^3}{M_R^5}+\mathcal{O}\left(\frac{M^3}{M_R^3}\frac{F^3}{M_R^5},\frac{F^5}{M_R^9}\right).
\end{equation}
On the other hand, the scalars masses are $\mathcal{O}(\alpha F/M_R)$. If R-symmetry is weakly broken $(M < M_R)$, a Mini-Split spectrum is again generated.

\subsection{Gaugino mass spectrum}
\begin{figure}
        \centering
        \begin{subfigure}[b]{0.475\textwidth}
                \centering
                \includegraphics[width=\textwidth,bb=0 0 260 206]{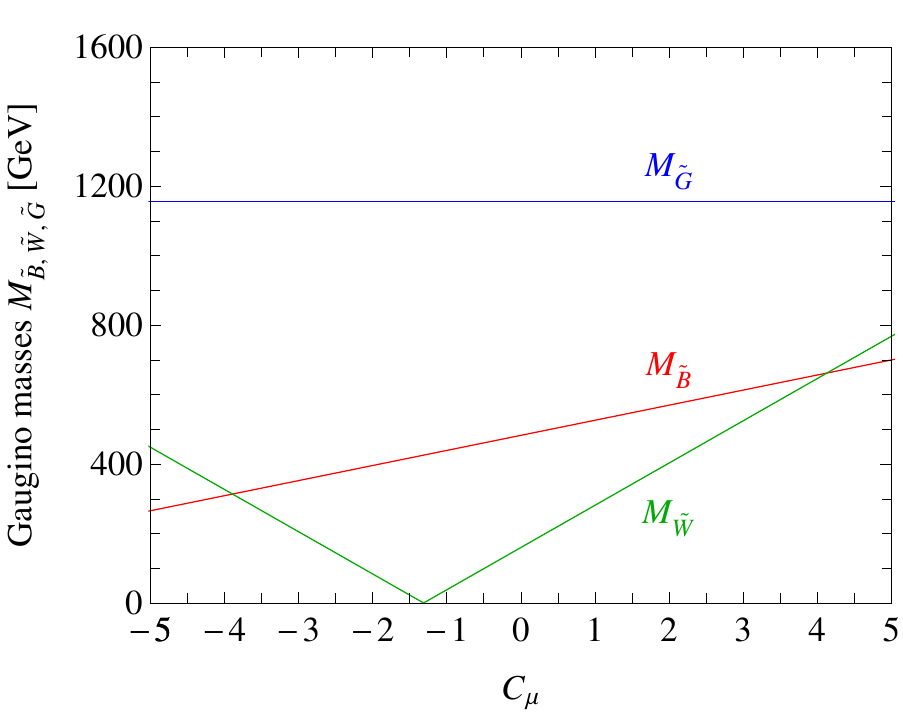}
                \caption{}
                \label{Figure1a}
        \end{subfigure}%
        ~\qquad
        \begin{subfigure}[b]{0.475\textwidth}
                \centering
                \includegraphics[width=\textwidth,bb=0 0 260 206]{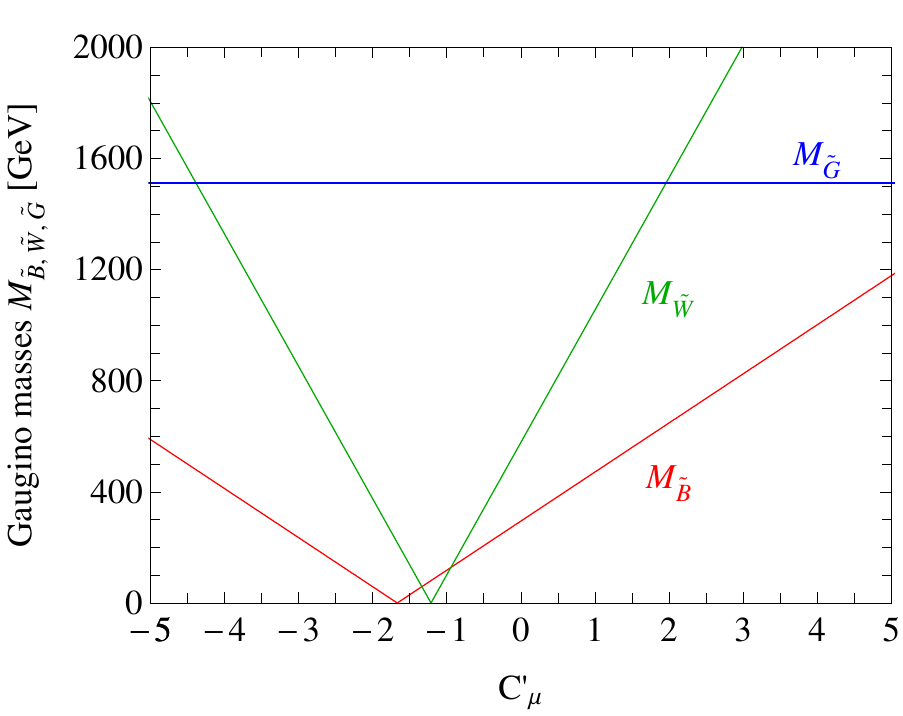}
                \caption{}
                \label{Figure1b}
        \end{subfigure}
				\caption{Typical mass spectrum for (a) anomaly mediation and (b) gauge mediation. In (a), the masses appearing on the right side of (\ref{AnomalyPoleMasses}) are taken to be $m_{\text{scalars}}=\mu=m_{3/2}=50$ TeV with $\tan\beta=2$. In (b), the masses appearing on the right side of (\ref{GaugePoleMasses}) are taken to be $m_{\text{scalars}}=\mu=\Lambda=200$ TeV with $\tan\beta=2$. }\label{PlotsMassSpectrum}
\end{figure}

The main effect of the small mass hierarchy between the gauginos and scalars/Higgsinos is that radiative corrections to the pole masses of gauginos coming from integrating out the scalars and Higgsinos  can be comparable to, if not larger than, the contributions coming from anomaly mediation or gauge mediation directly. In the case of anomaly mediation, the expressions are well known and can be read from different sources \cite{Bagnaschi:2014rsa, Gupta:2012gu}. In the limit of degenerate sfermion masses, the pole masses of the gauginos are
\begin{equation}\label{AnomalyPoleMasses}
   \begin{aligned}
	M_{\tilde{B}} = M_1(Q)&\left[1+\frac{C_\mu}{11}+\frac{8g_1^2}{80\pi^2}\left(-\frac{41}{2}\ln\frac{Q^2}{M_1^2}-\frac{1}{2}\ln\frac{\mu^2}{M_1^2}+\ln\frac{m_A^2}{M_1^2}\right.\right.\\
&\left.\left. +11\ln\frac{m_{\tilde{q}}^2}{M_1^2}+9\ln\frac{m_{\tilde{l}}^2}{M_1^2}\right)+\frac{g_3^2}{6\pi^2}-\frac{13g_t^2}{264\pi^2\sin^2\beta}\right]\\
	M_{\tilde{W}} = M_2(Q)&\left[1+C_\mu+\frac{g_2^2}{16\pi^2}\left(\frac{19}{6}\ln\frac{Q^2}{M_2^2}-\frac{1}{6}\ln\frac{\mu^2}{M_2^2}+\frac{1}{3}\ln\frac{m_A^2}{M_2^2}\right.\right.\\
&\left.\left. +3\ln\frac{m_{\tilde{q}}^2}{M_2^2}+\ln\frac{m_{\tilde{l}}^2}{M_2^2}\right)+\frac{3g_3^2}{2\pi^2}-\frac{3g_t^2}{8\pi^2\sin^2\beta}\right]\\
	M_{\tilde{G}} = M_3(Q)&\left[1+\frac{g_3^2}{16\pi^2}\left(7\ln\frac{Q^2}{M_3^2}+4\ln\frac{m_{\tilde{q}}^2}{M_3^2}+13-2F\left(\frac{M_3^2}{m_{\tilde{q}}^2} \right)  \right)-\frac{7g_3^2}{24\pi^2}+\frac{g_t^2}{12\pi^2\sin^2\beta}\right]
   \end{aligned}
\end{equation}
where
\begin{equation}\label{AnomalyBareMasses}
   M_1(Q) = \frac{33 g_1^2(Q)}{80\pi^2}m_{3/2} \;\;\;\; M_2(Q) = \frac{g_2^2(Q)}{16\pi^2}m_{3/2} \;\;\;\; M_3(Q) = -\frac{3g_3^2(Q)}{16\pi^2}m_{3/2},
\end{equation}
$g_i(Q)$ are the gauge couplings of the SM in $\overline{MS}$ and SU(5) convention at scale $Q$, $g_t$ is the top Yukawa coupling  in the SM, and
\begin{equation}
   \begin{aligned}
	C_\mu &= \frac{\mu}{m_{3/2}}\frac{m_A^2\sin^2\beta}{m_A^2-\mu^2}\ln\frac{m_A^2}{\mu^2}\\
	F(x)  &= 3\left[\frac{3}{2} -\frac{1}{x}-\left(\frac{1}{x}-1\right)^2\ln|1-x|\right].
   \end{aligned}
\end{equation}
The main point of interest is that the corrections due to $C_\mu$ can be comparable if not bigger than the usual expressions. A typical mass spectrum is shown in the left panel of figure \ref{PlotsMassSpectrum}. Similar expressions hold for gauge mediation
\begin{equation}\label{GaugePoleMasses}
   \begin{aligned}
	M_{\tilde{B}} = M'_1(Q)&\left[1+\frac{3C'_\mu}{5}+\frac{g_1^2}{80\pi^2}\left(-\frac{41}{2}\ln\frac{Q^2}{M_1^2}-\frac{1}{2}\ln\frac{\mu^2}{M_1^2}+\ln\frac{m_A^2}{M_1^2}+11\ln\frac{m_{\tilde{q}}^2}{M_1^2}+9\ln\frac{m_{\tilde{l}}^2}{M_1^2}\right)\right]\\
	M_{\tilde{W}} = M'_2(Q)&\left[1+C'_\mu+\frac{g_2^2}{16\pi^2}\left(\frac{19}{6}\ln\frac{Q^2}{M_2^2}-\frac{1}{6}\ln\frac{\mu^2}{M_2^2}+\frac{1}{3}\ln\frac{m_A^2}{M_2^2}+3\ln\frac{m_{\tilde{q}}^2}{M_2^2}+\ln\frac{m_{\tilde{l}}^2}{M_2^2}\right)\right]\\
	M_{\tilde{G}} = M'_3(Q)&\left[1+\frac{g_3^2}{16\pi^2}\left(7\ln\frac{Q^2}{M_3^2}+4\ln\frac{m_{\tilde{q}}^2}{M_3^2}+13-2F\left(\frac{M_3^2}{m_{\tilde{q}}^2} \right)  \right)+\frac{6g_3^2}{16\pi^2}\right]
   \end{aligned}
\end{equation}
where we have kept only the terms proportional to $g_t$, $g_3$, or log-enhanced \cite{Gherghetta:1999sw},
\begin{equation}
   \begin{aligned}
	&M'_i(Q) = \frac{g_i^2}{16\pi^2} \Lambda \\
	&C'_\mu =  \frac{\mu}{\Lambda} \frac{m_A^2\sin^2\beta}{m_A^2-\mu^2} \ln\frac{m_A^2}{\mu^2}
   \end{aligned}
\end{equation}
where $\Lambda$, in a given gauge mediation model, can be expressed in term of the SUSY breaking scale and the messenger scales (see for example eq. (\ref{eq:gaugemediation})). The last term of $M_{\tilde{G}}$ in (\ref{GaugePoleMasses}) can be extracted from \cite{Picariello:1998dy}. A typical mass spectrum is shown in the right panel of figure \ref{PlotsMassSpectrum}. 

The parameters $C_\mu$ and $C'_\mu$ can be rewritten by requiring the fine-tuning condition, which needs to be imposed to have the weak scale parametrically smaller than the scalars \cite{Bagnaschi:2014rsa},
\begin{equation}
   \tan^2\beta = \frac{m^2_{H_d}+\mu^2}{m^2_{H_u}+\mu^2}
\end{equation}
and the usual relation $m^2_A = m_{H_u}^2 + m_{H_d}^2 + 2\mu^2$. $C_{\mu}$ can then be expressed as \cite{Bagnaschi:2014rsa}
\begin{equation}\label{FinetuningCmu}
 C_\mu = \frac{2\mu\tan\beta}{m_{3/2}}\frac{m_{H_d}^2+\mu^2}{(\tan^2\beta+1)m_{H_d}^2+\mu^2}\ln\left[(1+\cot^2\beta)\left(1+\frac{m_{H_d}^2}{\mu^2}\right)\right].
\end{equation}
The same applies to $C'_\mu$ with $m_{3/2} \to \Lambda$.

In these models the gauginos are the lightest sparticles and, because $\mu$ is large, the light neutralinos and charginos are almost pure binos and winos. As such, there is a neutralino of mass very close to $M_{\tilde{B}}$ and a pair of nearly degenerate neutralino and chargino of mass $M_{\tilde{W}}$. There is a small mass difference between the neutral and charged wino dominated by a loop effect \cite{Feng:1999fu}
\begin{equation}\label{LoopSplitting}
   \Delta M \equiv m_{\chi_{\tilde{W}}^+}-m_{\chi_{\tilde{W}}^0} = \frac{\alpha_2 M_2}{4\pi}\left[f(r_W)-c_W^2f(r_Z)-s_W^2f(r_{\gamma})\right]
\end{equation}
where $f(y)=\int_0^1(2+2x)\log(x^2+(1-x)y^2)dx$ and $r_i=m_i/M_2$. The mass splitting is typically of the order of 150 MeV.

\subsection{Gaugino decays}\label{SectionGauginoDecays}
\begin{figure}[t!]
  \centering
  \includegraphics[width=0.5\textwidth,bb=0 0 360 315]{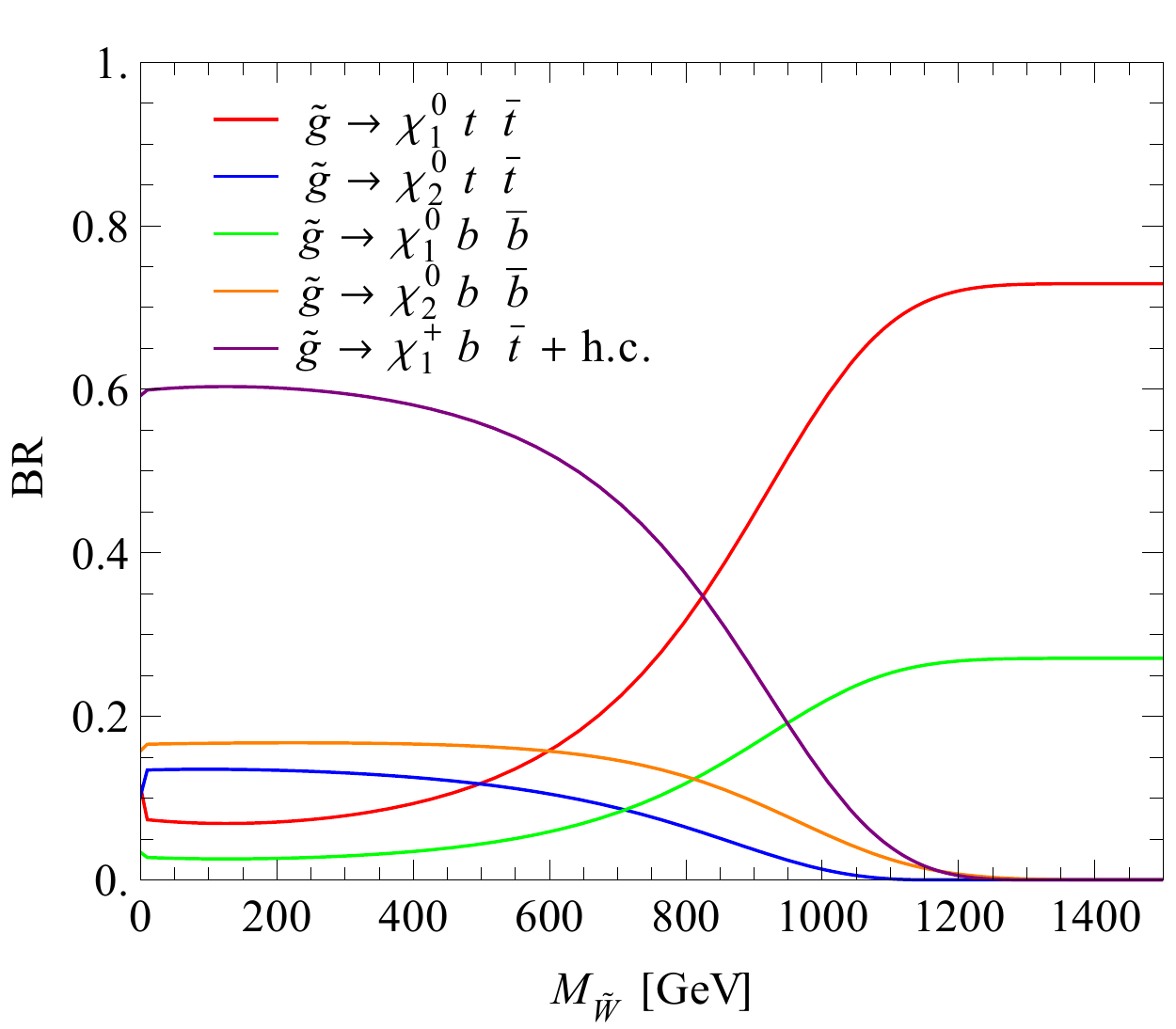}
  \caption{Branching ratios of the gluino for $M_{\tilde{G}}=1500$ GeV and $M_{\tilde{B}}=0$ GeV. The third generation scalar masses are assumed to be degenerate and much heavier than the gauginos.}\label{PlotBranching}
\end{figure}

In this work we concentrate on gluino decay via third generation squarks. These decay modes dominate if the third generation squarks are lighter than the others, which is expected from RG effects or could be imposed for other model building reasons.\footnote{For example, flavor physics might require the first and second generations of squarks to be in the 1000 TeV range, while the third generation could be kept somewhat lighter to obtain the appropriate Higgs mass \cite{ArkaniHamed:2012gw}.} The decays that we consider are then
\begin{equation}\label{eqGluinoDecay}
   \begin{aligned}
	&\tilde{g} \to t \overline{t} \chi_1^0\quad &&\tilde{g} \to b \overline{b} \chi_1^0\quad &&\tilde{g} \to b \overline{t} \chi_1^+ \\
	&\tilde{g} \to t \overline{t} \chi_2^0\quad &&\tilde{g} \to b \overline{b} \chi_2^0\quad &&\tilde{g} \to \overline{b} t \chi_1^-.
   \end{aligned}
\end{equation}
The gluino can also decay to a gluon and a neutralino; however, it is negligible for heavy enough Higgsinos \cite{Jung:2013zya} and we ignore it.  To compute the branching ratios we use analytical results that can be found in \cite{Toharia:2005gm}. An example of branching fractions is shown in figure \ref{PlotBranching}. In practice, $\chi_2^0$ always decays to $\chi_1^0$ and a Higgs boson \cite{ArkaniHamed:2012gw}, irrespective of whether $M_{\tilde{B}}$ is larger than $M_{\tilde{W}}$ or the opposite. In our scenario, the decay $\chi_2^0$ to $\chi_1^0$ and a Z boson is extremely suppressed due to the neutralinos being almost pure gauginos.  When $M_{\tilde{W}} < M_{\tilde{B}}$, $\chi_1^+$ can only decay to $\chi_1^0$ and either light leptons or a pion which can cause this chargino to be metastable because of lack of phase-space \cite{Feng:1999fu}. As the decay is always very soft, the decay products are generally unaccounted for and the chargino is practically indistinguishable from the stable neutralino. When $M_{\tilde{W}} > M_{\tilde{B}}$ , $\chi_1^+$ decays to $\chi_1^0$ and a $W$ boson (we verified that the decay to $\chi_2^0$ only becomes relevant for $\mu$ at a scale considerably higher than anything relevant to this paper). The branching ratios we compute assume equal masses for the stops and the sbottoms. If the stops were lighter, the decays to two $b$ quarks, which can only proceed via off-shell sbottoms, would be relatively suppressed. As can be seen in figure \ref{PlotBranching}, these decays are already suppressed. The only thing that would change is the branching fraction of $\tilde{g} \to t \overline{t} \chi_1^0$, $\tilde{g} \to t \overline{t} \chi_2^0$ and $\tilde{g} \to b \overline{t} \chi_1^+$, which all have similar efficiencies for the searches we consider. We therefore do not expect that this assumption will affect our results greatly.

\subsection{Higgs mass}\label{SectionHiggsMass}
To set the mass of the Higgs to its experimentally measured value, we follow the procedure outlined in \cite{Bagnaschi:2014rsa} which we summarize here. 
First, $\overline{MS}$ parameters are taken from reference \cite{Buttazzo:2013uya} for the top Yukawa and the gauge coupling constants and from \cite{Mihaila:2012pz} for the bottom and tau Yukawas. The quartic coupling of the Higgs boson is extracted from its pole mass \cite{Holthausen:2011aa, Giudice:2011cg, Sirlin:1985ux, Cabrera:2011bi} using a value of 125.15 GeV, which is the naive average of the ATLAS \cite{Aad:2014aba} and CMS \cite{Chatrchyan:2013mxa} values. These parameters are then evolved up to the scalars scale using three-loops beta functions \cite{Bednyakov:2012rb, Bednyakov:2012en, Bednyakov:2013eba}. Threshold corrections are taken from \cite{Bagnaschi:2014rsa}. These include one-loop corrections and two-loop QCD corrections. The Higgs quartic is then matched with its SUSY expression and the threshold corrections. This determines one of the parameters, therefore reducing the dimension of the parameter space by one. As explained in the next section,  we vary $\tan \beta$ to obtain the correct value of the Higgs mass.

In some regions of the parameter space it is not possible to obtain the correct Higgs mass because the required parameters lead to a color-breaking minimum that is deeper than the electroweak minimum. The necessary condition to avoid this is \cite{Bagnaschi:2014rsa}
\begin{equation}\label{eqVacuumStability}
   \frac{(A_t-\mu\cot\beta)^2}{m_{Q_3} m_{U_3}} < \left(4-\frac{1}{\sin^2\beta}\right)\left(\frac{m_{Q_3}^2}{m_{U_3}^2}+\frac{m_{U_3}^2}{m_{Q_3}^2}\right),
\end{equation}
where $m_{Q_3}$ is the third generation soft mass for the SU(2) quark doublet and $m_{U_3}$ the right-handed stop soft mass.

\section{Methodology and results}

\subsection{Parameter space}
We begin by discussing the parameter space we use to study the models of interest. It is very similar for both anomaly and gauge mediation. There are essentially four parameters that control the phenomenology of anomaly mediation \cite{Bagnaschi:2014rsa}. They are $m_{3/2}$, $\tan\beta$, $m_{\text{scalars}}$, and $\mu$. As explained in section \ref{SectionMiniSplitModels}, $m_{3/2}$ and $m_{\text{scalars}}$ are expected to be of the same order of magnitude so we set them equal to each other. An additional parameter can be fixed by requiring the theory to predict the correct mass of the Higgs boson with the help of the results of section \ref{SectionHiggsMass}. Generally speaking, $\tan\beta$ is the best parameter to do so as varying it even slightly can have a substantial effect on the Higgs mass. The parameter space is then reduced to $\mu$ and  $m_{3/2}$. However, we trade $\mu$ for $C_\mu$. The main advantage of this parametrization is that the ratio of gaugino masses depends mostly on $C_\mu$. The exact details of the scalar sector are relegated to two-loops corrections in (\ref{AnomalyPoleMasses}) and our results can therefore be applied to models where the scalar sector does not differ too significantly.  To translate this to something more familiar, we provide each parameter space plot with contours of constant $M_{\tilde{B}}$, $M_{\tilde{W}}$, $\mu$, and $\tan\beta$.  

The relationship between $\mu$ and $C_\mu$ depends on $m_{H_d}$ which we take to be at $m_{\text{scalars}}$.  A different choice would lead, for the same $C_\mu$, to a different value of $\mu$ which in turn would affect mostly the color-breaking bounds (see equation (\ref{eqVacuumStability})). Taking $m_{H_d}$ much bigger than $m_{\text{scalars}}$ would limit $C_{\mu}$ to a narrow band around 0 and taking $m_{H_d}$ much smaller would push the bounds to large values of $C_\mu$ such that the gluino would be the LSP for most of the parameter space. With $m_{H_d}$ being set to $m_{\text{scalars}}$, we have a benchmark that does not suffer from any of these drawbacks. We assume the third generation to be lighter than the others, so as a benchmark we set the first and second generation squarks masses to $4m_{\text{scalars}}$ and all third generation masses to $m_{\text{scalars}}$. This is small enough to prevent problems with large logs, while keeping branching fraction to the first two generations below the percent level which is well below some of the uncertainties (e.g. gluino pair production cross-section). Sleptons masses are also set to $4m_{\text{scalars}}$.  Lowering the masses of the first two generations of squarks would increase the branching ratio of the gluino to light jets, possibly affecting the reach of our searches (however, the high jet-multiplicity would still provide strong bounds). It would have only a slight effect on the gaugino spectrum and on the Higgs mass. Finally, we set  the third generation A-term $A_t$ by equation (\ref{eqAterm}). Overall, changing our choice of benchmark parameters (mainly the choice of setting $m_{H_d}$ to $m_{\text{scalars}}$ and  of taking $m_{Q_3}=m_{U_3}=m_{\text{scalars}}$)   will mostly affect the $\mu$ and $\tan \beta$ contours in our results. Also, as a result of a modified relationship between $C_{\mu}$, $\mu$, and $\tan \beta$, the region of parameter space where there is a color-breaking vacuum would also be modified. 

In almost all of our parameter space the Higgsinos are heavy, except for a region near $C_\mu=0$ where a Higgsino can be the lightest superpartner (LSP). More precisely, outside of $|C_\mu|<0.3$, the Higgsinos are always an order of magnitude heavier than the gluino while only inside $|C_\mu|<0.1$ are the Higgsinos comparable in mass to the bino and winos. This represents only a very narrow band in the parameter space and the efficiencies of the signal regions are not expected to change much in it. In addition, this case has already been studied in \cite{Jung:2013zya, Barnard:2012au}. As such, we neglect this effect. When $M_{\tilde{W}} < M_{\tilde{B}}$, the mass difference between $\chi_1^+$ and $\chi_1^0$ is calculated using (\ref{LoopSplitting}). 

The previous discussion applies almost directly to gauge mediation by trading $m_{3/2}$ for $\Lambda$. In this case, we fix $m_{\text{scalars}}$ to $\Lambda$ while $\tan \beta$ is again set by requesting the correct mass of the Higgs boson.\footnote{There is considerable freedom on the choice of the scalar masses. The choice we make is more to keep in tune with our procedure for anomaly mediation. As explained above, the exact details of the scalar sector are not very relevant in our parametrization.} The masses of the sleptons and the first two generations squarks are still set to $4m_{\text{scalars}}$. $A_t$ is set to zero, as one would expect it to be small \cite{Bagnaschi:2014rsa}, and is then completely overshadowed by $\mu$. The mass $m_{H_d}$ is once more set to $m_{\text{scalars}}$. 

Two other constraints are of importance for the parameter space. First of all, for a given value of $m_{3/2}$ ($\Lambda$), a small value of $A_t$ will lead to an upper bound on $C_\mu$ ($C'_\mu$)  beyond which it is impossible to obtain the correct Higgs mass. Indeed if $C_\mu$ ($C'_\mu$) becomes large, the threshold corrections also become large and the quartic matching condition does not accept any solutions for real $\tan\beta$. In fact, requiring $C_\mu$ ($C'_\mu$) close to its upper bound can make the Higgsinos heavy enough that large logs could become a problem and perturbation expansions could fail. Fixing the stop mixing parameter $A_t-\mu\cot\beta$ to a small value would solve this problem, but this would imply $A_t$ reaching values that are too high to be readily explained in our framework without large fine-tuning. The second issue arises from the presence of a color-breaking vacuum which is controlled by equation (\ref{eqVacuumStability}). For the values of $m_{3/2}$ ($\Lambda$) considered in this work, it turns out that this limit is always stronger than the upper bound on $C_\mu$ ($C'_\mu$) coming from the mass of the Higgs boson. This latter constraint can therefore be ignored. We limit ourselves to the regions of parameter space where equation (\ref{eqVacuumStability}) is satisfied.

\subsection{Current LHC constraints}
To obtain current limits on anomaly and gauge mediation, we recast  searches for gluino pair production. In particular  we concentrate on searches with either many b-jets, leptons, or large jet-multiplicity. Of course, all of these searches have stringent cuts on missing transverse energy (MET). The chosen searches are summarized in table \ref{TableSearches}. As a general rule, \cite{Aad:2014lra} dominates over the others. For each of these searches, we implemented codes simulating the cuts. To validate our codes, we generated events with MadGraph 5 \cite{Alwall:2014hca} intefaced with Pythia 6 \cite{Sjostrand:2006za} and Delphes 3 \cite{deFavereau:2013fsa, Cacciari:2011ma}. We were able to reproduce all four searches with good accuracy. There are also constraints coming from electrowino production for which the experimental bounds found in \cite{TheATLAScollaboration:2013zia, Aad:2014vma, Aad:2014nua, Khachatryan:2014mma} apply directly. This is because the branching ratios for the charginos and neutralinos that are relevant for our models are the same as the one used in the simplified models considered in those searches. The bounds are in general much weaker than the one from gluino production and become relevant only in a tiny region of parameter space where the electrowinos are very light.

\begin{table}[t!]
  \begin{center}
	\begin{tabular}{ |C{3cm}|C{4cm}|C{4cm}|C{3cm}| }
	\hline
  Collaboration	& Search							& Strategy																									& Reference 										\\ \hline
	ATLAS					& JHEP 06 (2014) 035	& 2 same sign / 3 leptons + 0-3 b-jets + MET								& \cite{Aad:2014pda}						\\ \hline
	ATLAS					& JHEP 10 (2014) 024	& 0-1 leptons + ≥ 3 b-jets + MET														& \cite{Aad:2014lra}						\\ \hline
	CMS						& CMS-SUS-13-012			& High jet-multiplicity + MET																& \cite{Chatrchyan:2014lfa}		  \\ \hline
	CMS						& CMS-PAS-SUS-12-016	& 2 opposite sign leptons + high-jet multiplicity + ≥3 b-jets + MET	& \cite{CMS:2013ija}		\\ \hline
	\end{tabular}%
	\caption{Gluino pair production searches.}\label{TableSearches}
	\end{center}
\end{table}

Our method to reinterpret the experimental constraints follows closely the procedure of \cite{Papucci:2014rja}. We look at every possible combination of decay chains (\ref{eqGluinoDecay}) and evaluate for each of them the efficiency of every signal region. The branching fractions are then calculated using the procedure of section \ref{SectionGauginoDecays}. The gluino pair production cross-sections are calculated at NLO+NLL with NLL-fast \cite{Beenakker:1996ch, Kulesza:2008jb, Kulesza:2009kq, Beenakker:2009ha, Beenakker:2011fu}, which we verified using Prospino \cite{Beenakker:1996ed}. The number of expected signals in a given signal region can then be calculated. The $95\%$ confidence level signal upper limit can either be read directly from these searches or calculated using the known background and confidence level (CL) techniques \cite{Read:2002hq}. The different signal regions are combined in a boolean fashion \cite{Martin:2014qra}. A more thorough approach would require the correlation between the backgrounds of the different signal regions, which is not readily available.

The events are generated with MadGraph 5 \cite{Alwall:2014hca} interfaced with Pythia 6 \cite{Sjostrand:2006za} and Delphes 3 \cite{deFavereau:2013fsa, Cacciari:2011ma}. 10000 events are generated for each grid point. MadGraph generally takes care of decay chains up to the production of the LSP. The only exception is when either $\chi_2^0$ or $\chi_1^+$ are very close in mass to $\chi_1^0$. These decays can then be forced to be off-shell and the decay chains become too long to be handled by MadGraph comfortably. In the worst case scenario, $\chi_2^0$ can decay to $\chi_1^0$ and an off-shell Higgs which then decays to a W and a off-shell W which in turn decays to other particles. To handle these difficult decays, we calculate branching ratios in advance using the decay functionalities of MadGraph to produce decay tables. $\chi_2^0$ and $\chi_1^+$ are then decayed  by Pythia using these results.  Delphes handles the detector simulation and is tuned to simulate the ATLAS and CMS detectors.

\begin{figure}
        \centering
        \begin{subfigure}[b]{0.5\textwidth}
                \centering
                \caption{$M_{\tilde{B}}$ [GeV]}
                \includegraphics[width=\textwidth,bb=0 0 580 390]{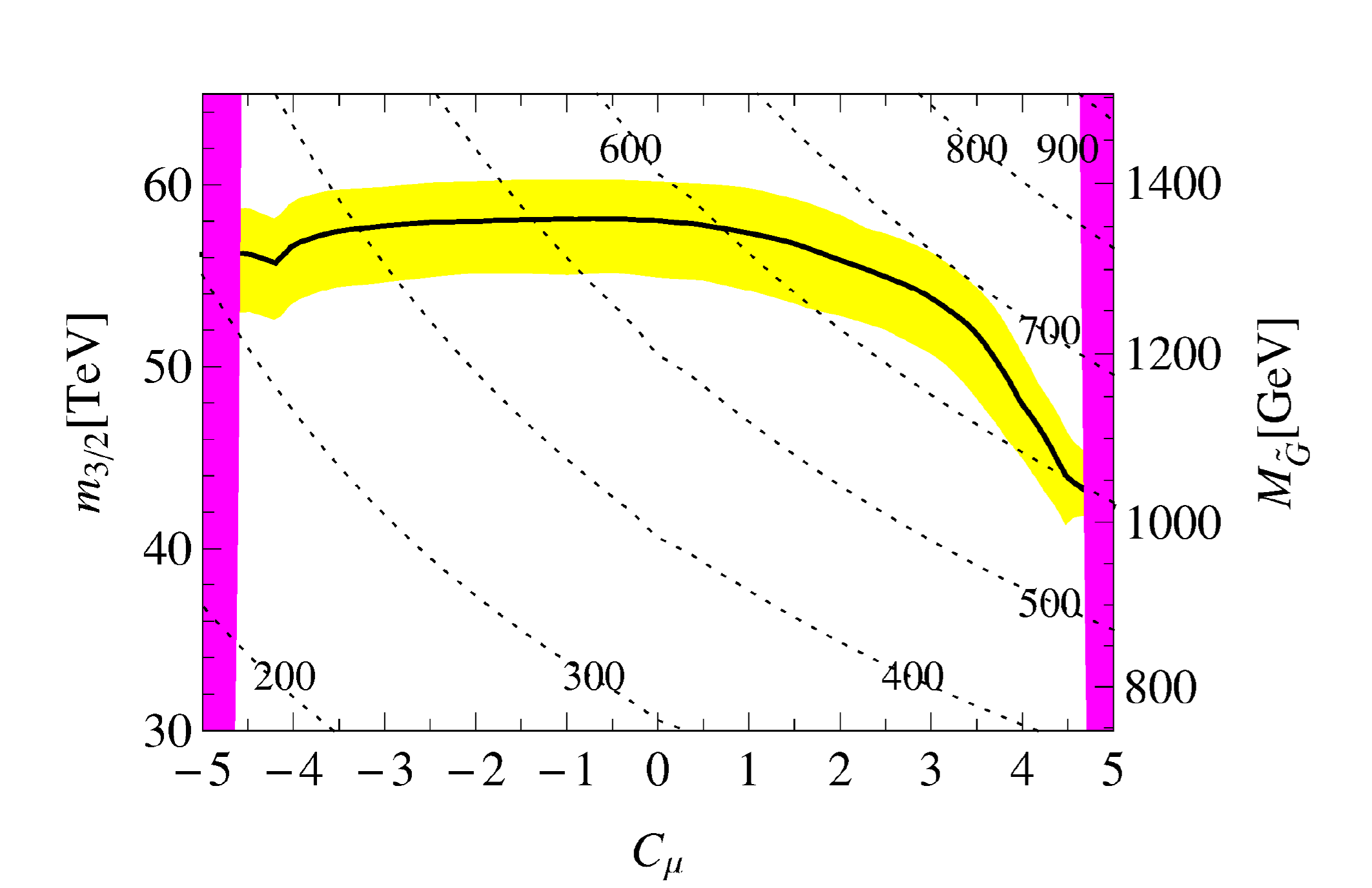}
                \label{Plot1a}
        \end{subfigure}%
        ~
        \begin{subfigure}[b]{0.5\textwidth}
                \centering
                \caption{$M_{\tilde{W}}$ [GeV]}      
                \includegraphics[width=\textwidth,bb=0 0 580 390]{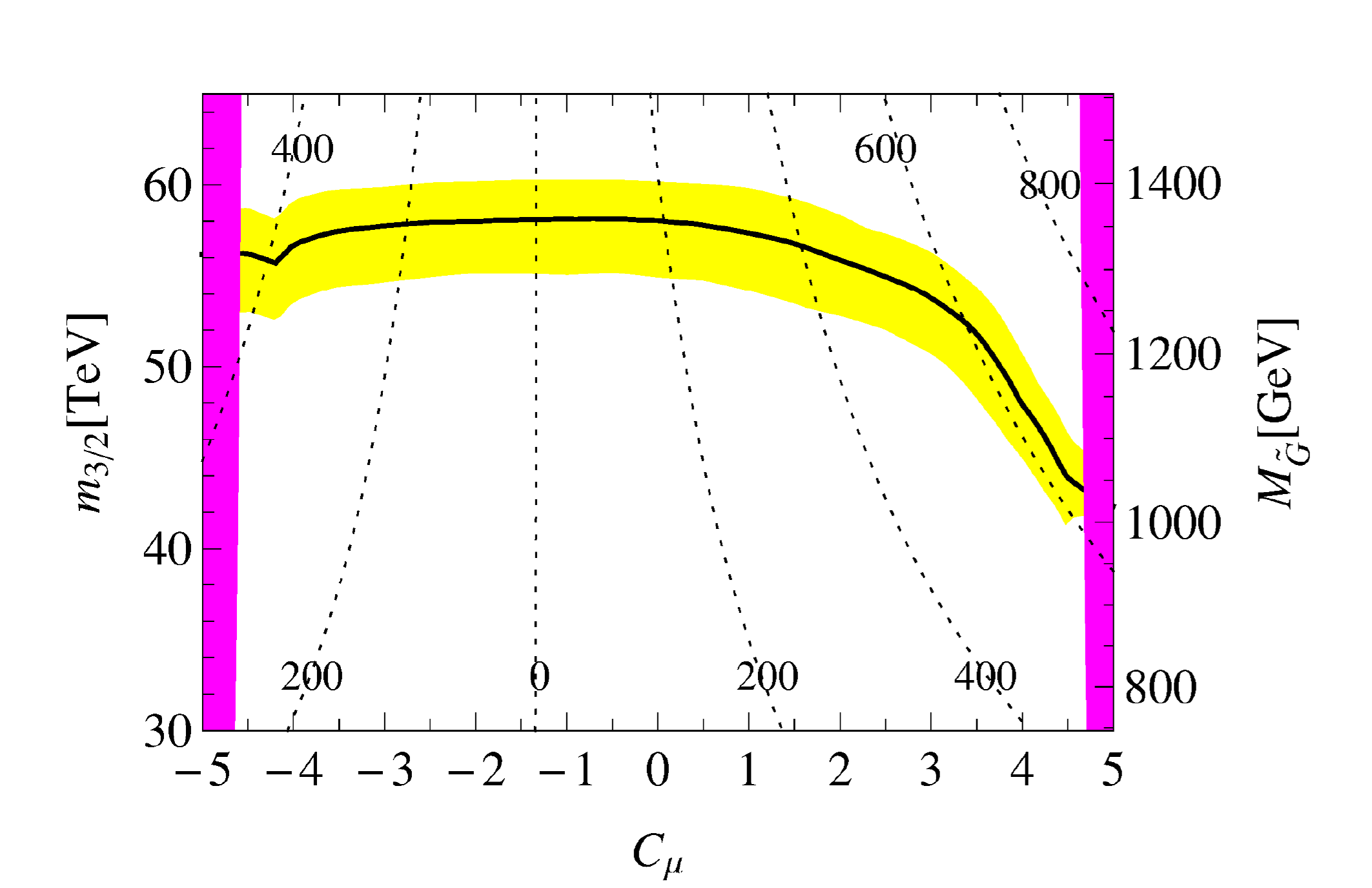}
                \label{Plot1b}
        \end{subfigure}
        \begin{subfigure}[b]{0.5\textwidth}
                \centering
                \caption{$\mu$ [TeV]}
                \includegraphics[width=\textwidth,bb=0 0 580 390]{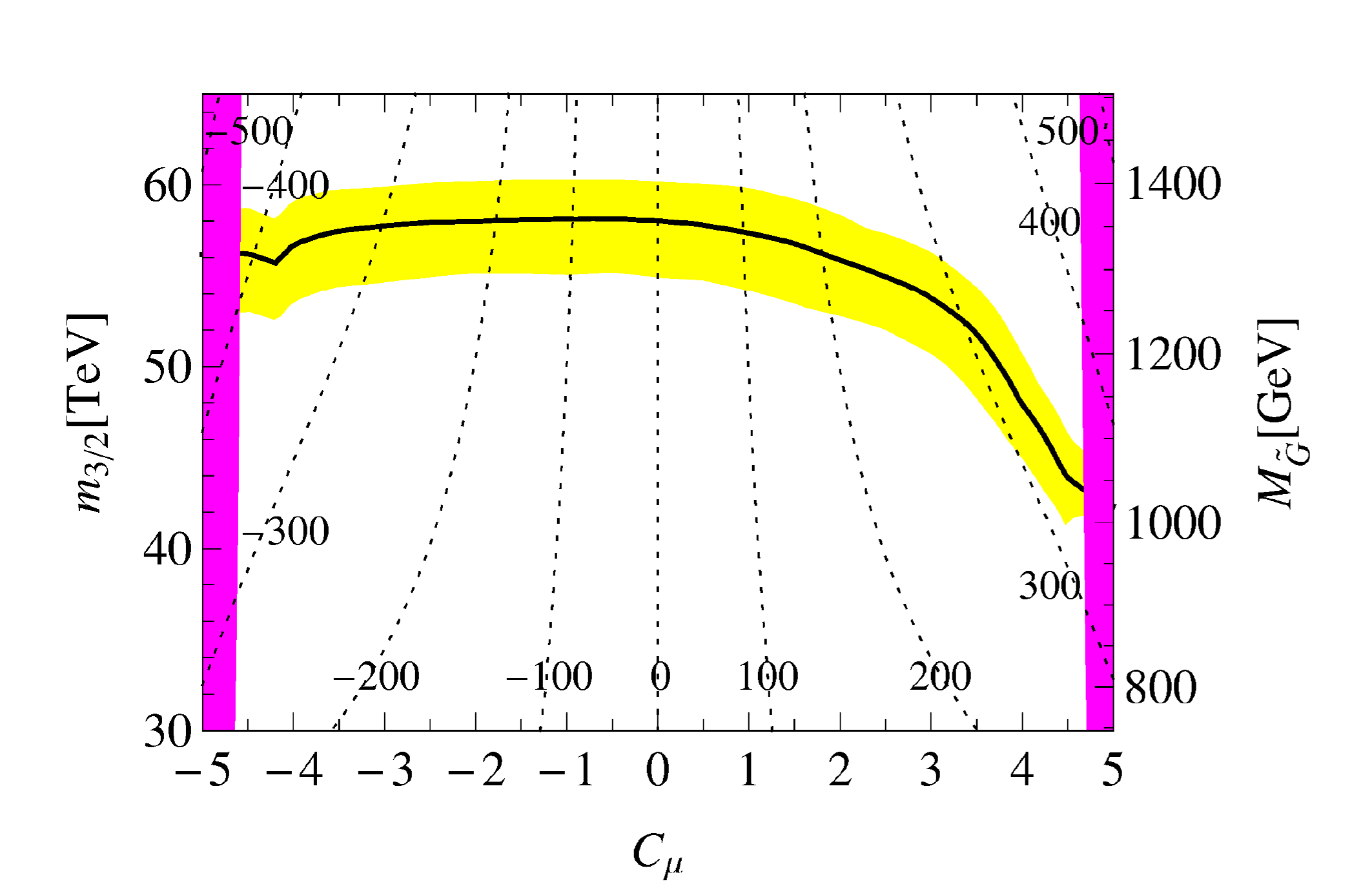}
                \label{Plot1c}
        \end{subfigure}%
        ~
        \begin{subfigure}[b]{0.5\textwidth}
                \centering
                \caption{$\tan\beta$}      
                \includegraphics[width=\textwidth,bb=0 0 580 390]{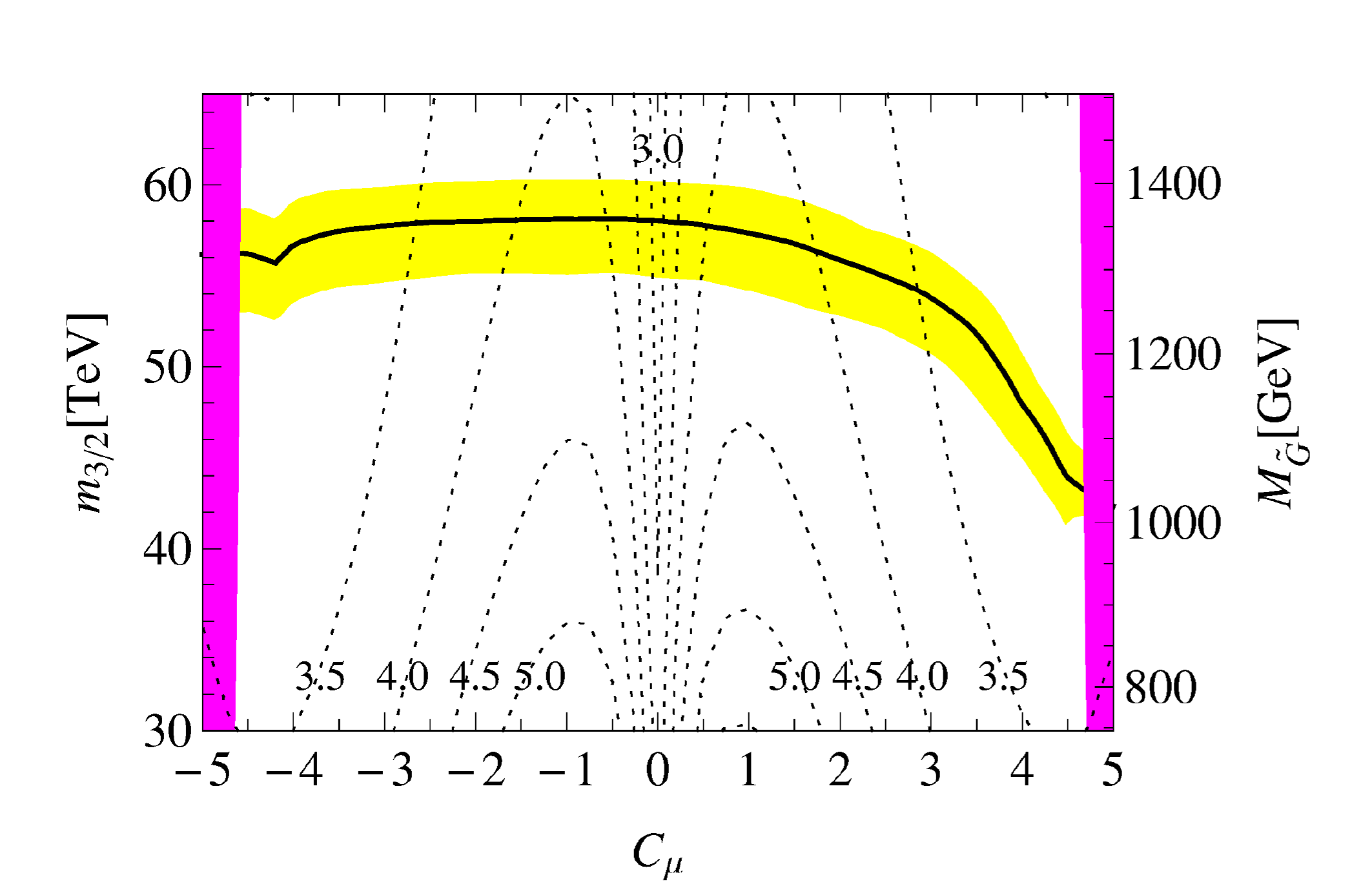}
                \label{Plot1d}
        \end{subfigure}
				\caption{95$\%$ exclusion limits for anomaly mediation at the LHC. The yellow band corresponds to the $1\sigma$ uncertainty on the gluino pair production cross-section and the purple bands are the forbidden region of color-breaking vacuum. Contour lines of constant $M_{\tilde{B}}$, $M_{\tilde{W}}$, $\mu$, and $\tan\beta$ are shown respectively in (a), (b), (c), and (d).}\label{AnomalyLHC}
\end{figure}

\begin{figure}
        \centering
        \begin{subfigure}[b]{0.5\textwidth}
                \centering
                \caption{$M_{\tilde{B}}$ [GeV]}
                \includegraphics[width=\textwidth,bb=0 0 580 390]{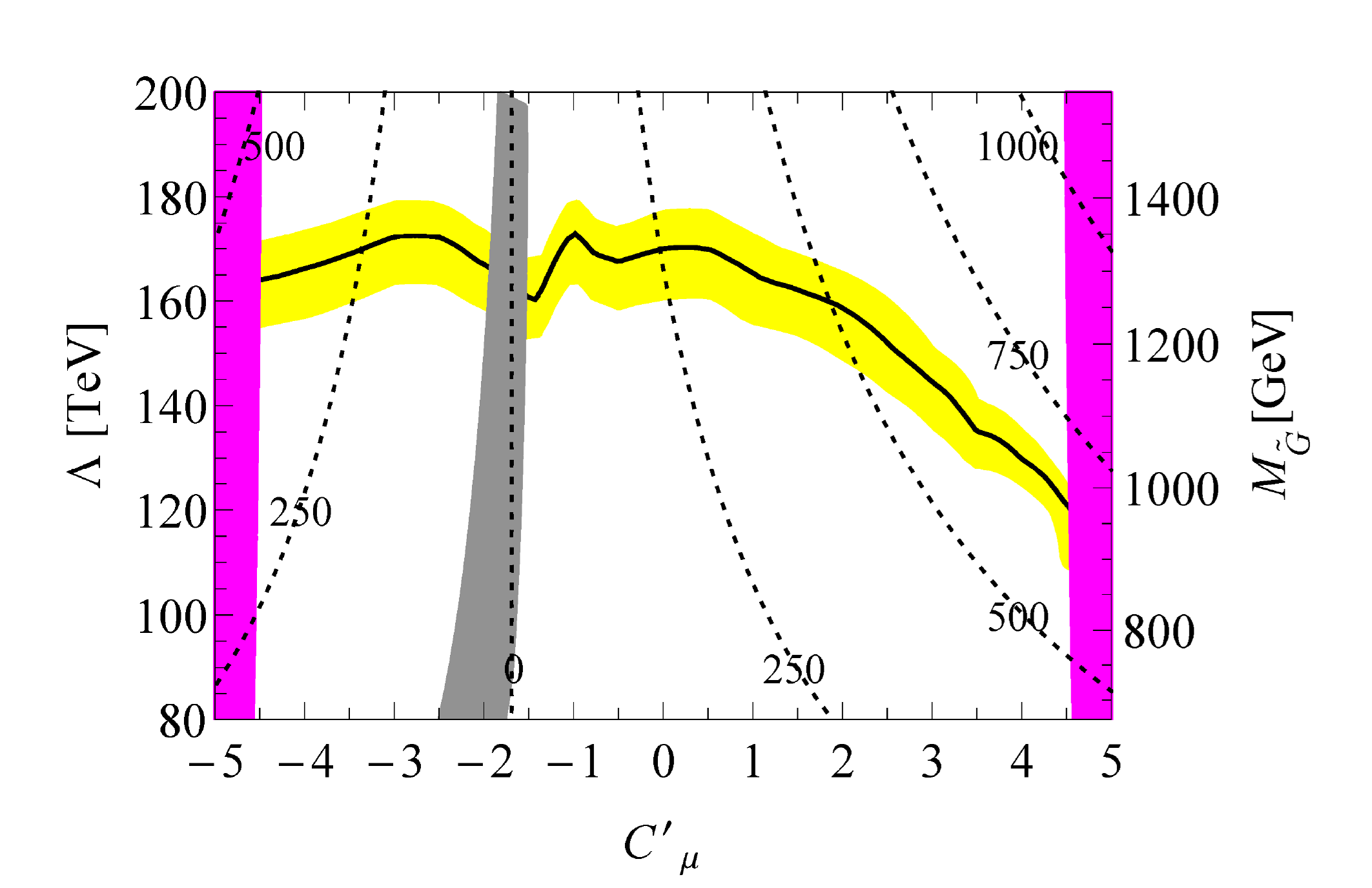}
                \label{Plot2a}
        \end{subfigure}%
        ~
        \begin{subfigure}[b]{0.5\textwidth}
                \centering
                \caption{$M_{\tilde{W}}$ [GeV]}      
                \includegraphics[width=\textwidth,bb=0 0 580 390]{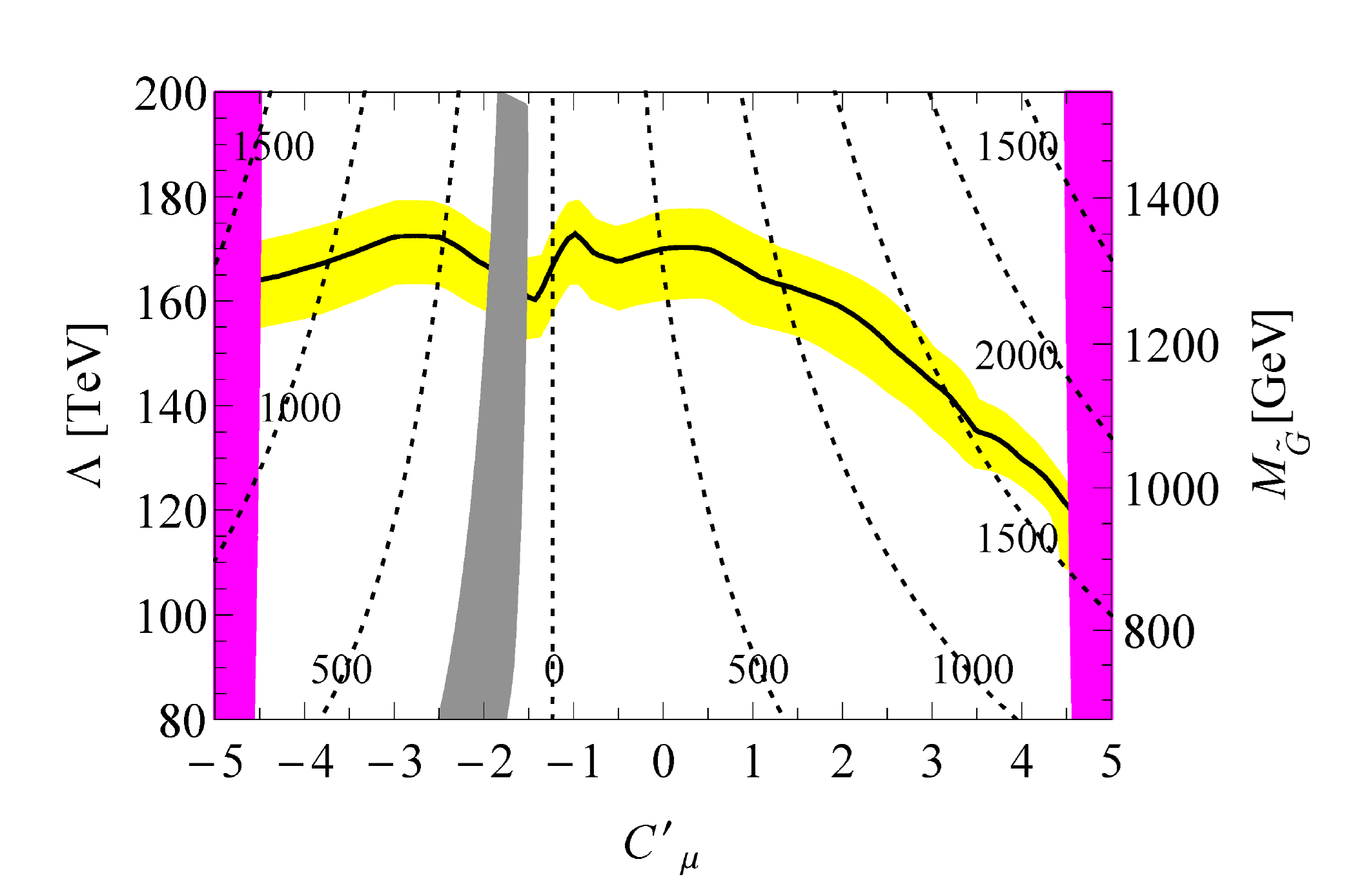}
                \label{Plot2b}
        \end{subfigure}
        \begin{subfigure}[b]{0.5\textwidth}
                \centering
                \caption{$\mu$ [TeV]}
                \includegraphics[width=\textwidth,bb=0 0 580 390]{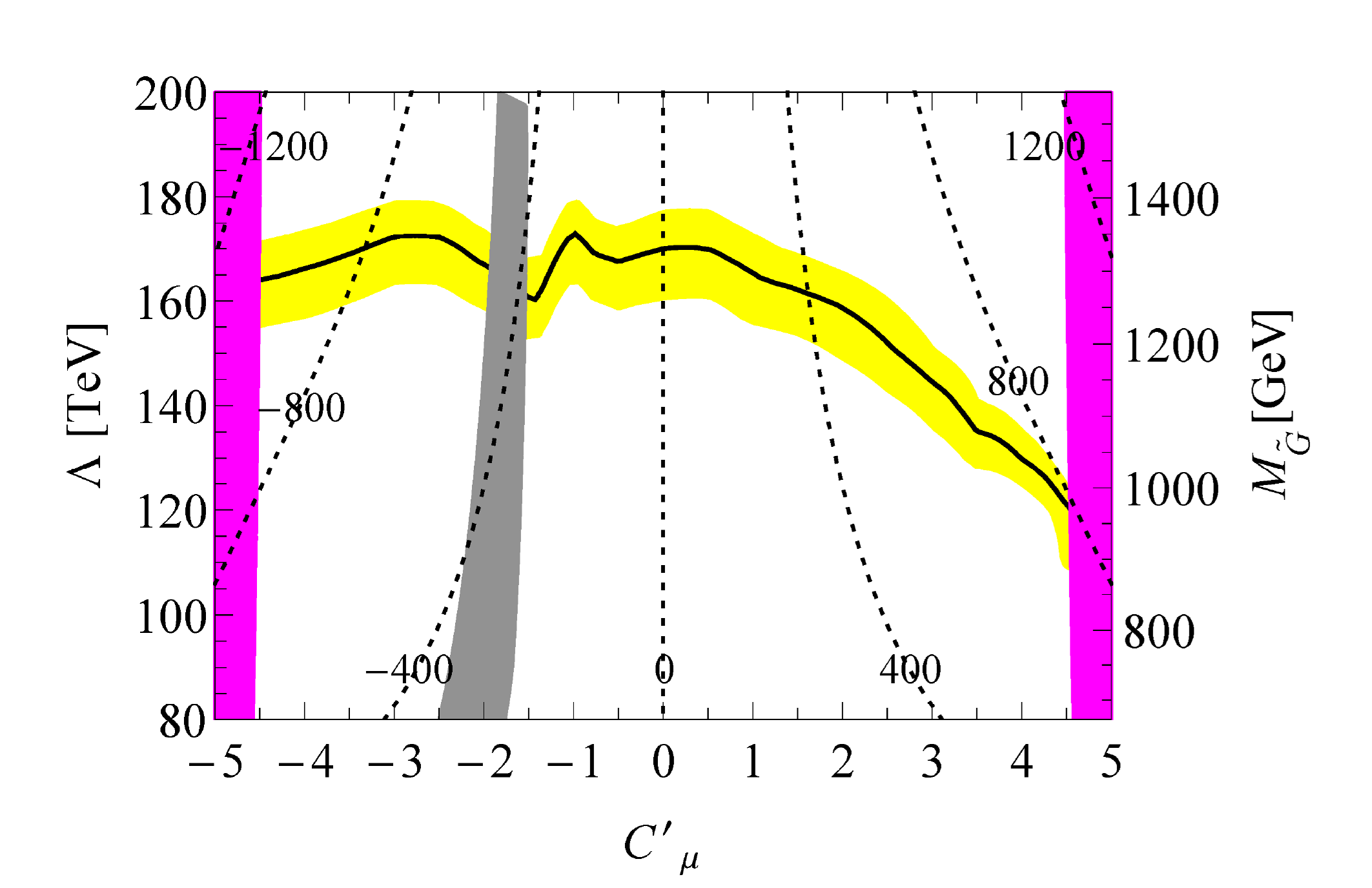}
                \label{Plot2c}
        \end{subfigure}%
        ~
        \begin{subfigure}[b]{0.5\textwidth}
                \centering
                \caption{$\tan\beta$}      
                \includegraphics[width=\textwidth,bb=0 0 580 390]{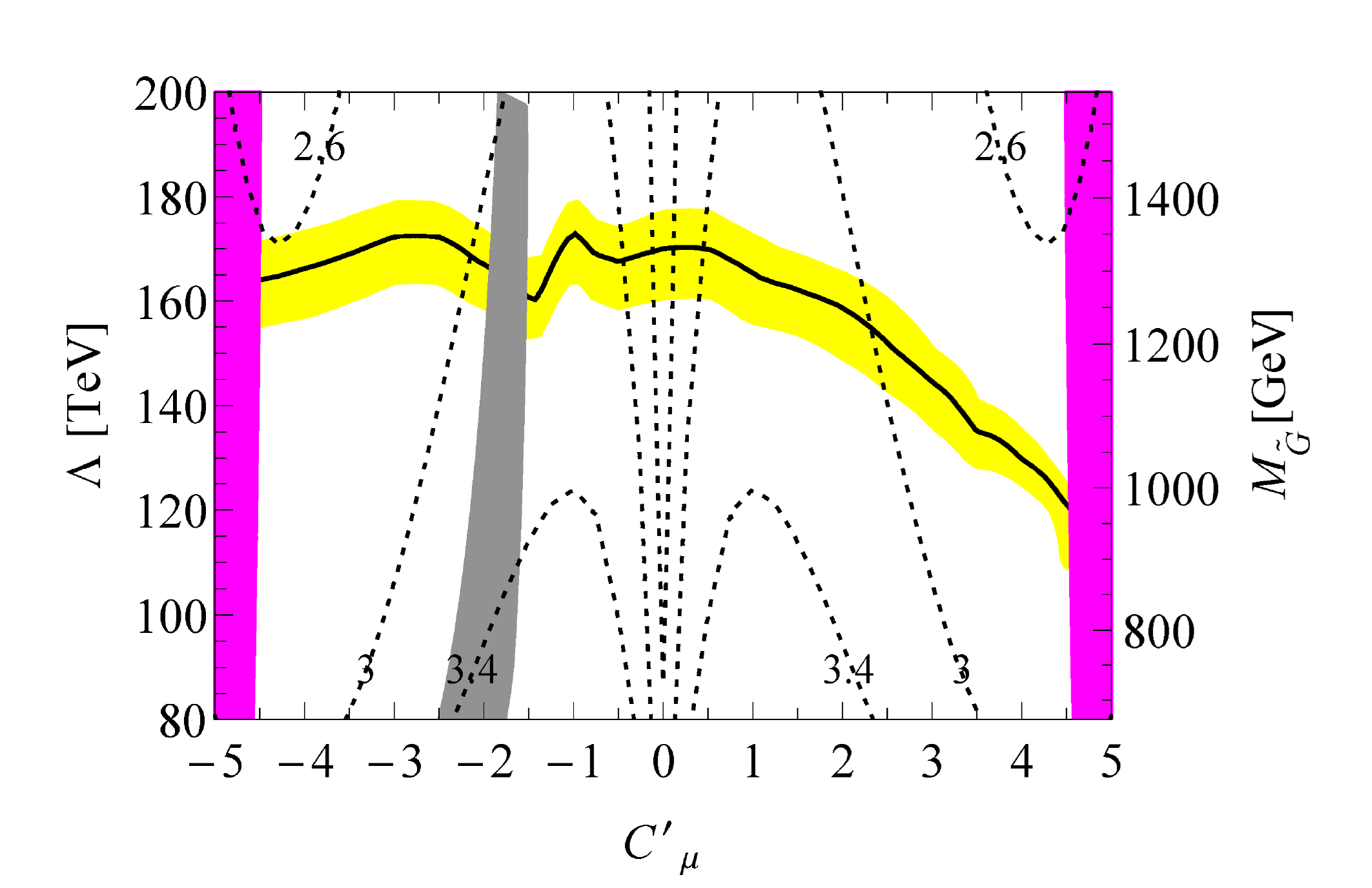}
                \label{Plot2d}
        \end{subfigure}
				\caption{95$\%$ exclusion limits for gauge mediation at the LHC. The yellow band corresponds to the $1\sigma$ uncertainty on the gluino pair production cross-section and the purple bands are the forbidden region of color-breaking vacuum. The grey band corresponds to limits from direct electroweak searches. Contour lines of constant $M_{\tilde{B}}$, $M_{\tilde{W}}$, $\mu$, and $\tan\beta$ are shown respectively in (a), (b), (c), and (d).}\label{GaugeLHC}
\end{figure}

The results for the 95$\%$ CL limits from ATLAS and CMS are given in figures \ref{AnomalyLHC} and \ref{GaugeLHC} for anomaly and gauge mediation respectively. Each one is provided with contour plots of $M_{\tilde{B}}$, $M_{\tilde{W}}$, $\mu$, and $\tan\beta$ to relate it to more familiar parameters. The regions forbidden by color-breaking vacuum are shown in purple. Overall, gluinos of mass up to $1.3$ TeV can be excluded over significant regions of parameter space. The results for the anomaly mediation spectrum can be easily understood. Over the entire covered parameter space, the gluino decays mainly to charginos. For $C_\mu$ between -4 and 4, the neutral wino is the LSP. The most relevant parameter in this region is then the ratio of the mass of the LSP and of the gluino. Below $C_\mu$ equal to 2, this ratio is large and the exclusion limits are strong. Above that value, the mass spectrum becomes compressed and kinematics quantities like MET become much smaller. As such, the exclusion limits drop considerably.

The results for gauge mediation are similar but with a few additional subtleties. Near $C'_\mu$ equal to -5, the spectrum is fairly compressed and the wino is too heavy to be produced. The gluino decays softly to $\chi_1^0$ and quarks, which results in lower constraints. As $C'_\mu$ increases, the spectrum becomes less compressed and the limits are stronger. However, near $C'_\mu$ equal to -3, the winos become light enough to be produced and the gluino decay to chargino dominates. As these decay chains are longer, there is less MET and the constraints are less strong. In a very narrow band around $C'_\mu$ equal -1.5, the wino is the LSP. The chargino then decays softly to a neutral wino. This is similar to gluino decaying to $\chi_1^0$ and the exclusion reaches the same levels as at $C'_\mu$ equal to -3. As $C'_\mu$ continues to increase, the mass spectrum again becomes compressed to the point where gluinos can only decay to $\chi_1^0$ and a pair of soft bottom quarks and the limits drop considerably. In addition, direct electroweakinos production searches from \cite{TheATLAScollaboration:2013zia, Aad:2014vma, Aad:2014nua, Khachatryan:2014mma} impose limits in a very narrow band near $C'_\mu$ equal to -2. This corresponds to when both the wino and bino are light which only occurs around $C'_\mu$ equal to $-2$. This region is shown as a grey band in figure \ref{GaugeLHC}. 

\subsection{Prospects at LHC 14}\label{SectionProspects14}
The procedure of the previous section can be modified to predict the discovery and exclusion prospects at the next phase of the LHC. The only differences amount to the signal regions and background estimations.

Two different strategies are adopted to cover the possibilities of the spectrum being compressed or not. When the LSP is considerably lighter than the gluino, kinematic quantities like MET are large and strong kinematic cuts are sufficient to eliminate most of the background. We refer to these signal regions as high MET cuts. On the contrary, when the gluino has a mass close to the LSP, quantities like MET become small and the cuts remove most signals. Lowering the cuts does not improve the limits much as the background increases considerably. However, adding the requirement of a pair of same sign dilepton (SSDL) drastically cuts the background and allows the kinematic cuts to be made less stringent by exploiting the possible production of leptons during the top decay. The only drawback to SSDL is that a large part of the signal is cut and the resulting limits are less strong than pure high MET cuts in the non-compact case. The net result is that high MET signal regions usually dominate until the spectrum becomes near degenerate. The exclusion then drops until the signal regions with SSDL become relevant which prevents the exclusion limits from dropping too fast. However, the SSDL cuts eventually also fail when there is not enough phase space for the gluino to produce top quarks.

For the high MET signal regions, we adopt the cuts of \cite{CMS:2013ega} for gluino decaying to top quarks and a single lepton. The cuts for SSDL are taken directly from \cite{Cohen:2013xda} and correspond to their gluino-neutralino model with heavy flavor decay for 14 TeV. We verified that we could reproduce both sets of results.

The detector card for Delphes is the standard 14 TeV card from Snowmass \cite{Anderson:2013kxz}. The background estimates for the high MET regions are obtained from the Snowmass online backgrounds \cite{Avetisyan:2013onh}. We simply apply our cuts on their events while taking into consideration their relative weight. The Snowmass backgrounds also provide events files with different average number of pile-up. In general, pile-up has very little effect on the high MET regions, while, for SSDL, leptons can possibly get lost in the pile-up jets \cite{Cohen:2013xda}, reducing the efficiency of the signal. We however concentrate on the case of 0 pile-up as the effect is generally small on most of the parameter space. For high MET cuts, we obtain backgrounds of (23.0, 12.1, 2.6, 2.1) for the four signal regions of \cite{CMS:2013ega} and 3000 $\text{fb}^{-1}$ of integrated luminosity. This can be compared with their result at 140 pile-up of (17.5, 4.8, 0.9, 1.6) and the same integrated luminosity. The backgrounds for SSDL are taken directly from \cite{Cohen:2013xda}, as we follow very closely their procedure. A 20$\%$ systematic uncertainty on all backgrounds is assumed \cite{Cohen:2013xda}. The gluino pair production cross-section is calculated using NLL-fast \cite{Beenakker:1996ch, Kulesza:2008jb, Kulesza:2009kq, Beenakker:2009ha, Beenakker:2011fu} customized for a 14 TeV collider. The possibility of 300 and 3000 $\text{fb}^{-1}$ of integrated luminosity are considered.

\begin{figure}
        \centering
        \begin{subfigure}[b]{0.5\textwidth}
                \centering
                \caption{$M_{\tilde{B}}$ [TeV]}
                \includegraphics[width=\textwidth,bb=0 0 400 275]{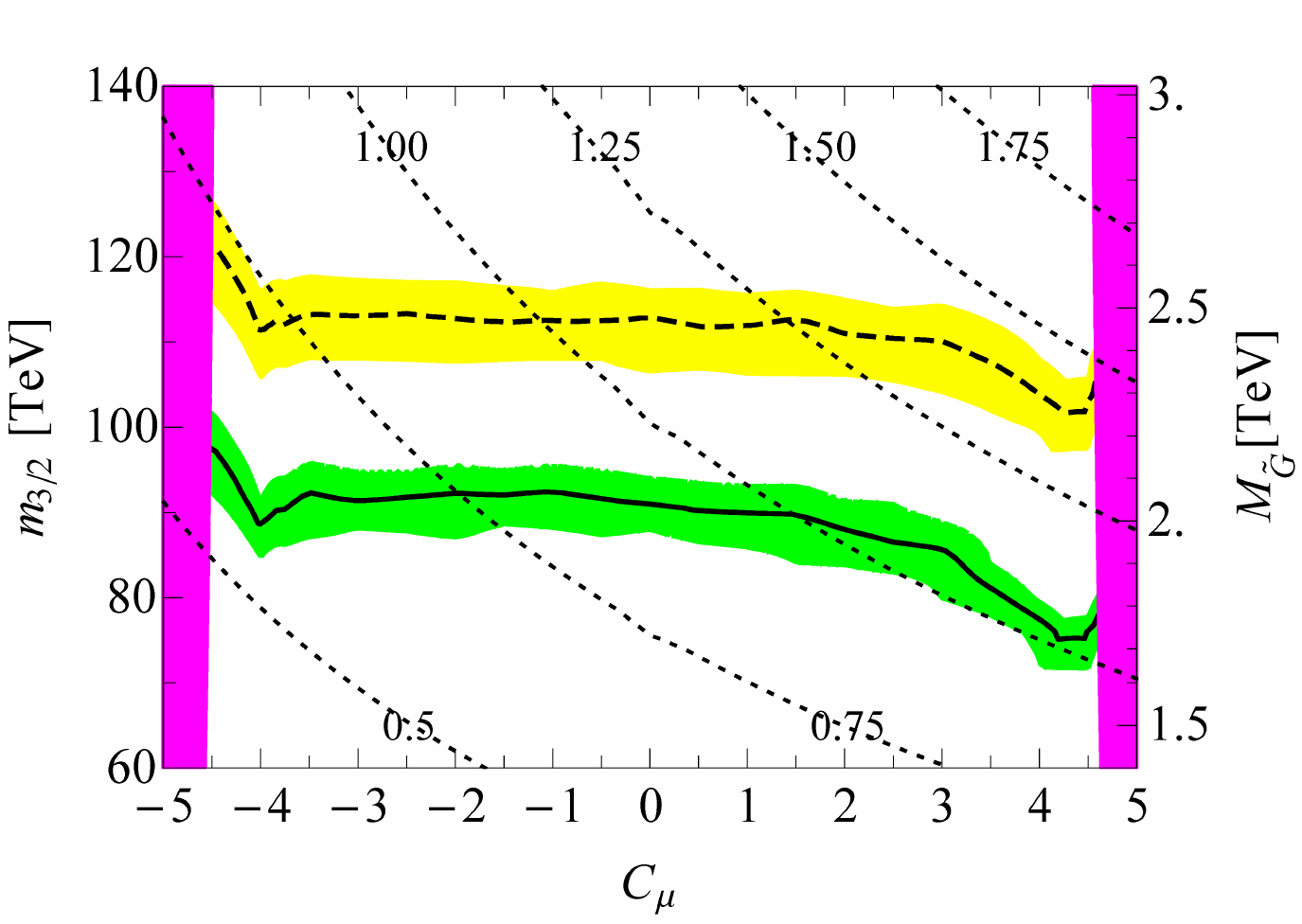}
                \label{Plot2a}
        \end{subfigure}%
        ~
        \begin{subfigure}[b]{0.5\textwidth}
                \centering
                \caption{$M_{\tilde{W}}$ [TeV]}      
                \includegraphics[width=\textwidth,bb=0 0 400 275]{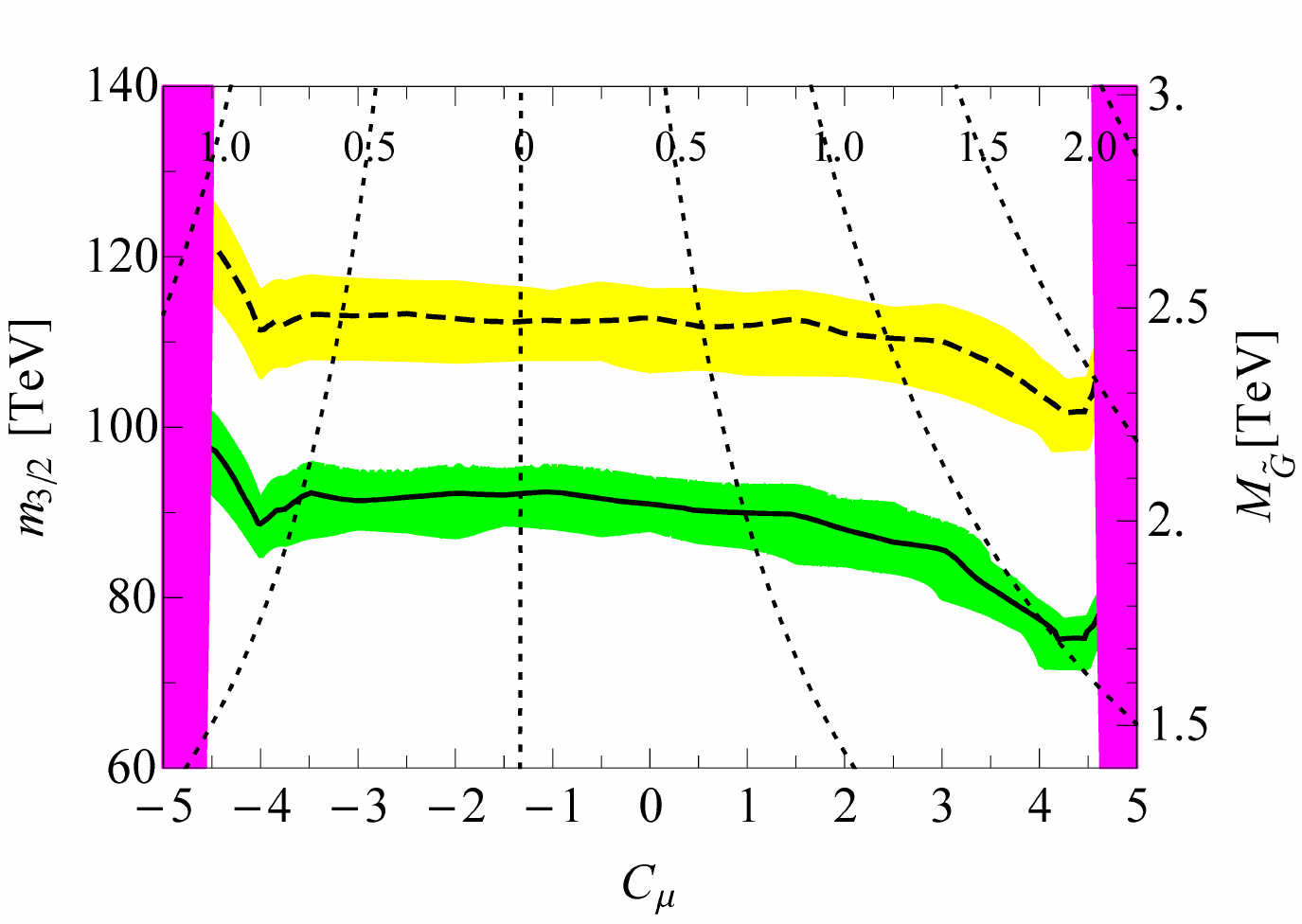}
                \label{Plot2b}
        \end{subfigure}
        \begin{subfigure}[b]{0.5\textwidth}
                \centering
                \caption{$\mu$ [PeV]}
                \includegraphics[width=\textwidth,bb=0 0 400 275]{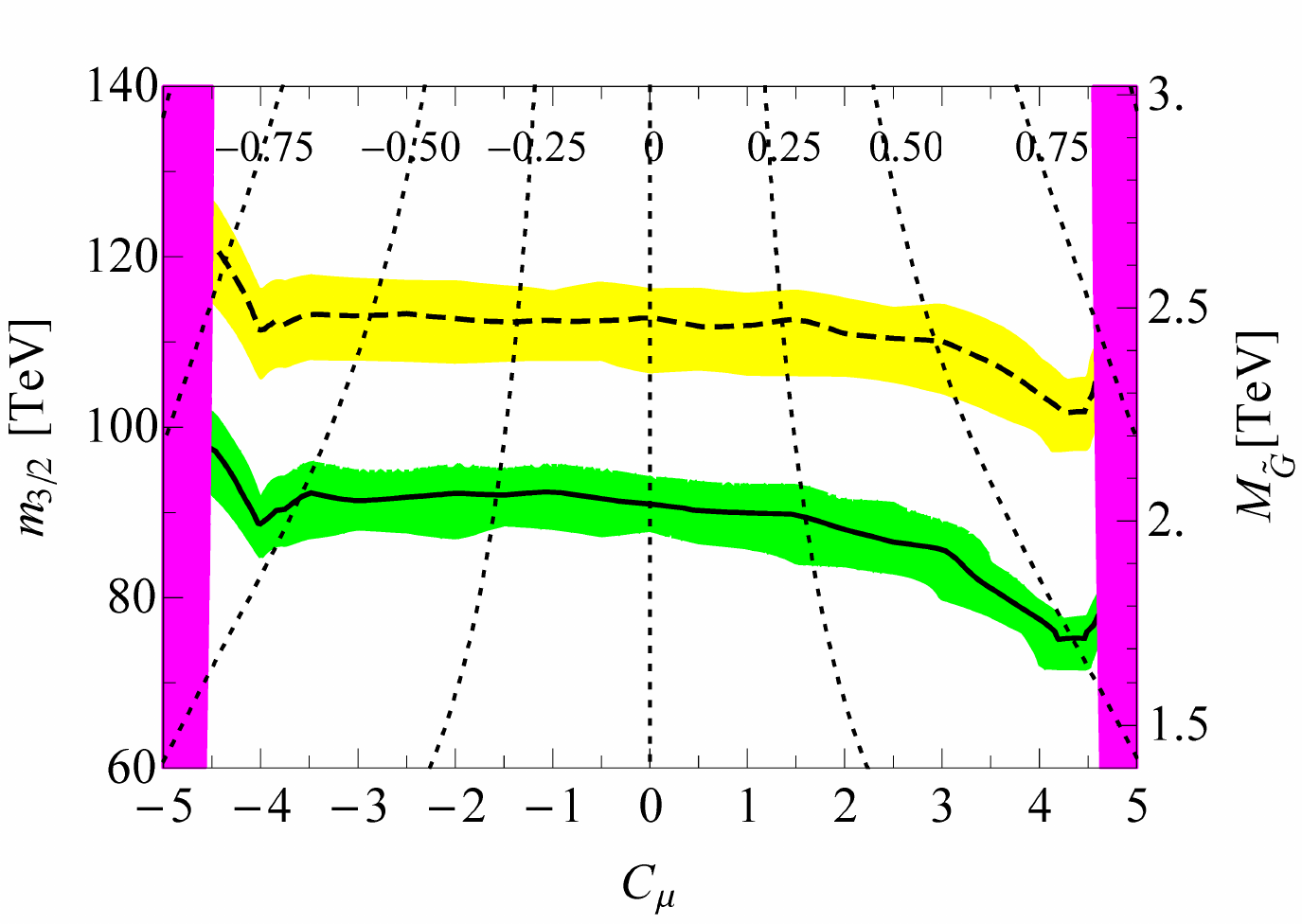}
                \label{Plot2c}
        \end{subfigure}%
        ~
        \begin{subfigure}[b]{0.5\textwidth}
                \centering
                \caption{$\tan\beta$}      
                \includegraphics[width=\textwidth,bb=0 0 400 275]{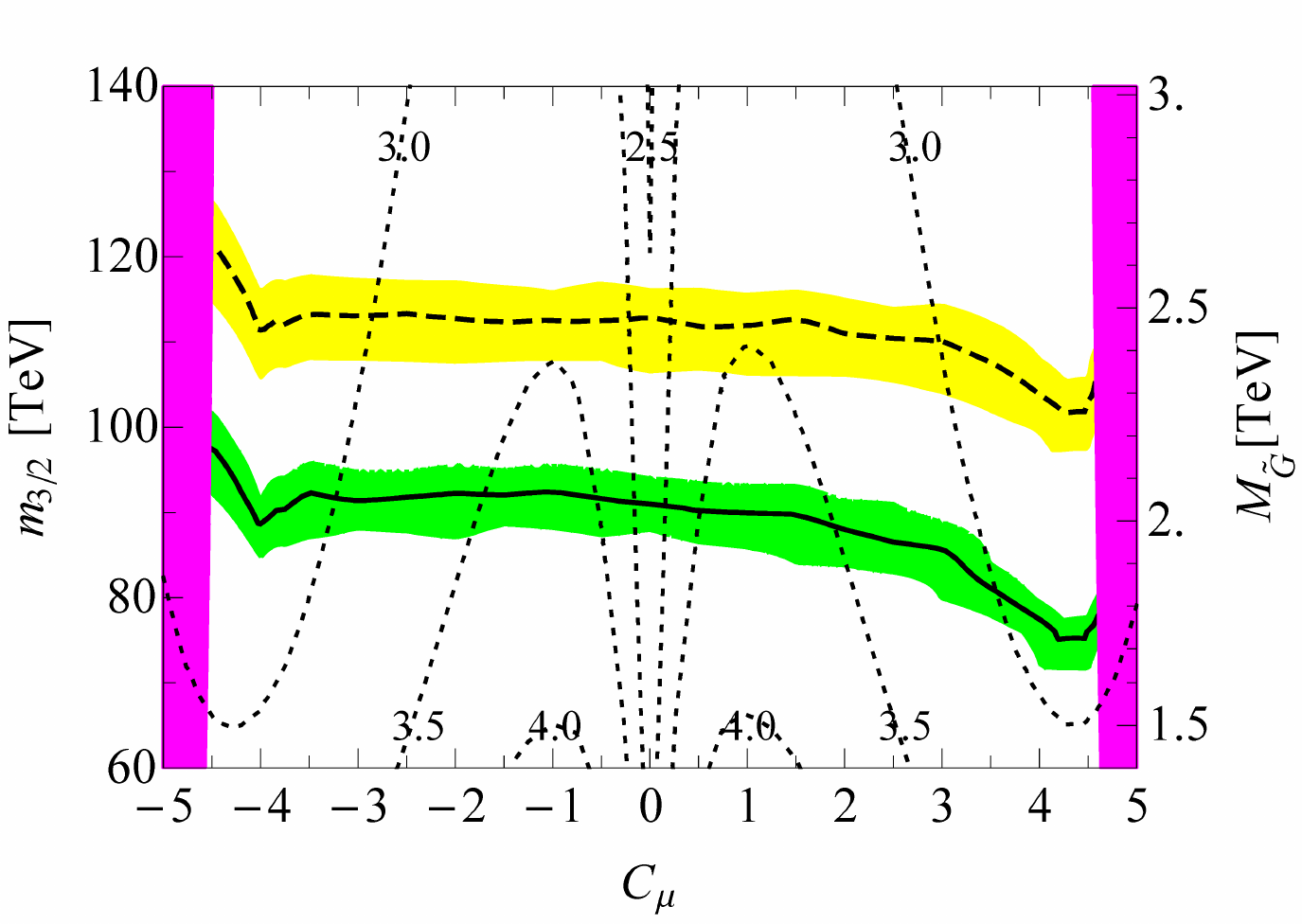}
                \label{Plot2d}
        \end{subfigure}
				\caption{95$\%$ exclusion limits for anomaly mediation at LHC 14 for (solid) 300 $\text{fb}^{-1}$ and (dashed) 3000 $\text{fb}^{-1}$ integrated luminosity. The green band corresponds to the $1\sigma$ uncertainty on the gluino pair production cross-section for 300 $\text{fb}^{-1}$, the yellow band corresponds to the $1\sigma$ uncertainty on the gluino pair production cross-section for 3000 $\text{fb}^{-1}$, and the purple bands are the forbidden region of color-breaking vacuum. Contour lines of constant $M_{\tilde{B}}$, $M_{\tilde{W}}$, $\mu$, and $\tan\beta$ are shown respectively in (a), (b), (c), and (d).}\label{AnomalyLHC1495}
\end{figure}

\begin{figure}
        \centering
        \begin{subfigure}[b]{0.5\textwidth}
                \centering
                \caption{$M_{\tilde{B}}$ [TeV]}
                \includegraphics[width=\textwidth,bb=0 0 400 275]{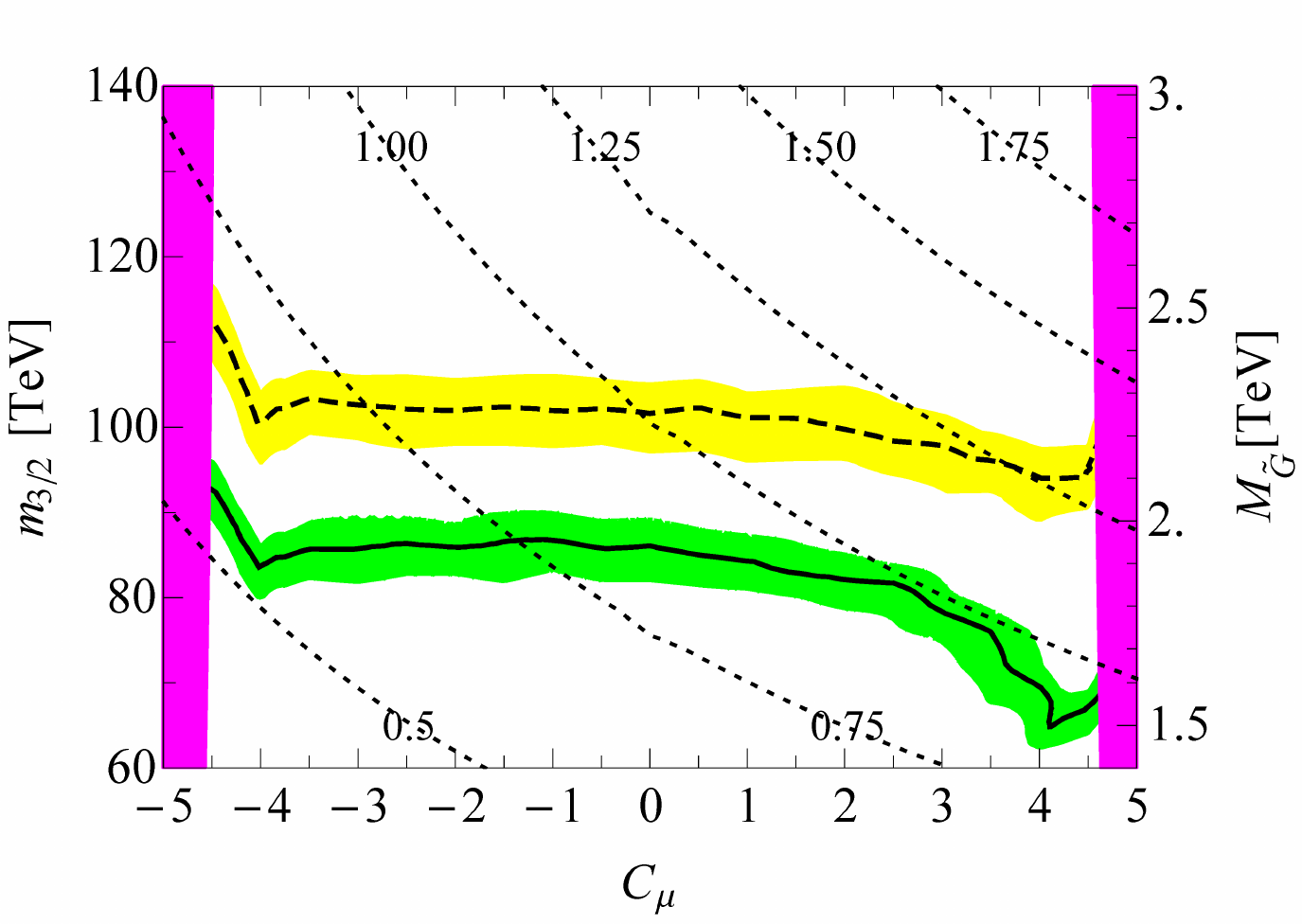}
                \label{Plot2a}
        \end{subfigure}%
        ~
        \begin{subfigure}[b]{0.5\textwidth}
                \centering
                \caption{$M_{\tilde{W}}$ [TeV]}      
                \includegraphics[width=\textwidth,bb=0 0 400 275]{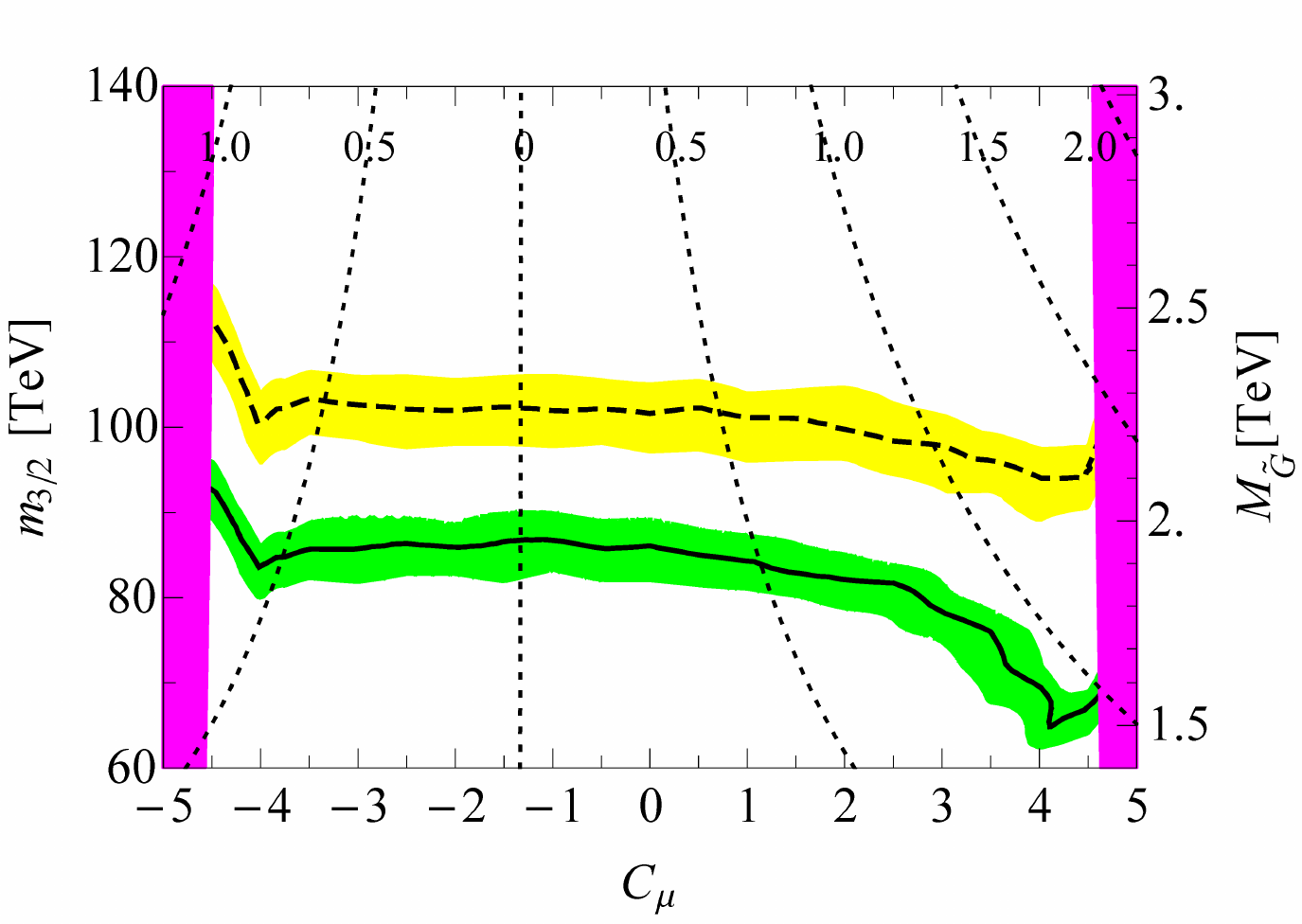}
                \label{Plot2b}
        \end{subfigure}
        \begin{subfigure}[b]{0.5\textwidth}
                \centering
                \caption{$\mu$ [PeV]}
                \includegraphics[width=\textwidth,bb=0 0 400 275]{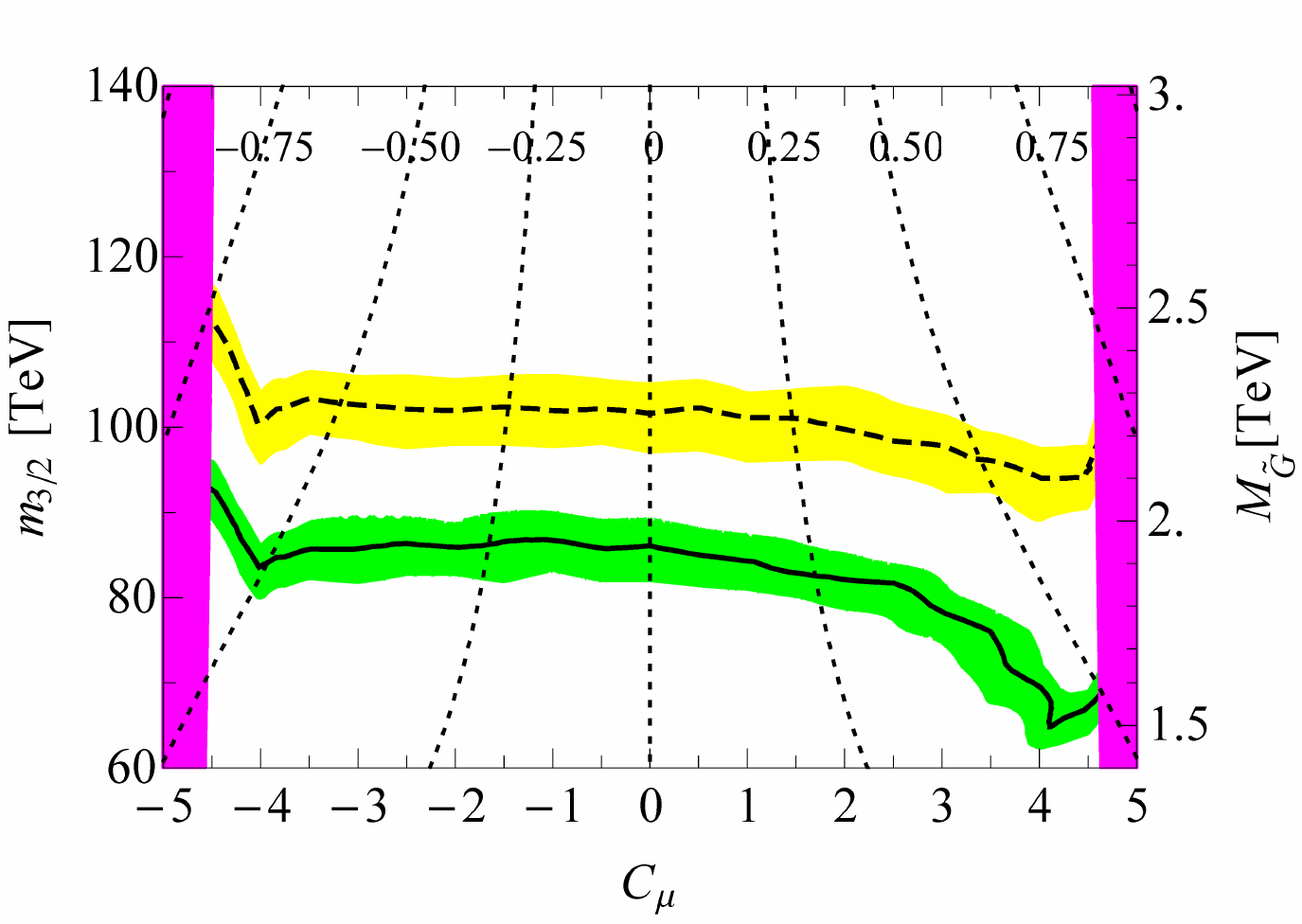}
                \label{Plot2c}
        \end{subfigure}%
        ~
        \begin{subfigure}[b]{0.5\textwidth}
                \centering
                \caption{$\tan\beta$}      
                \includegraphics[width=\textwidth,bb=0 0 400 275]{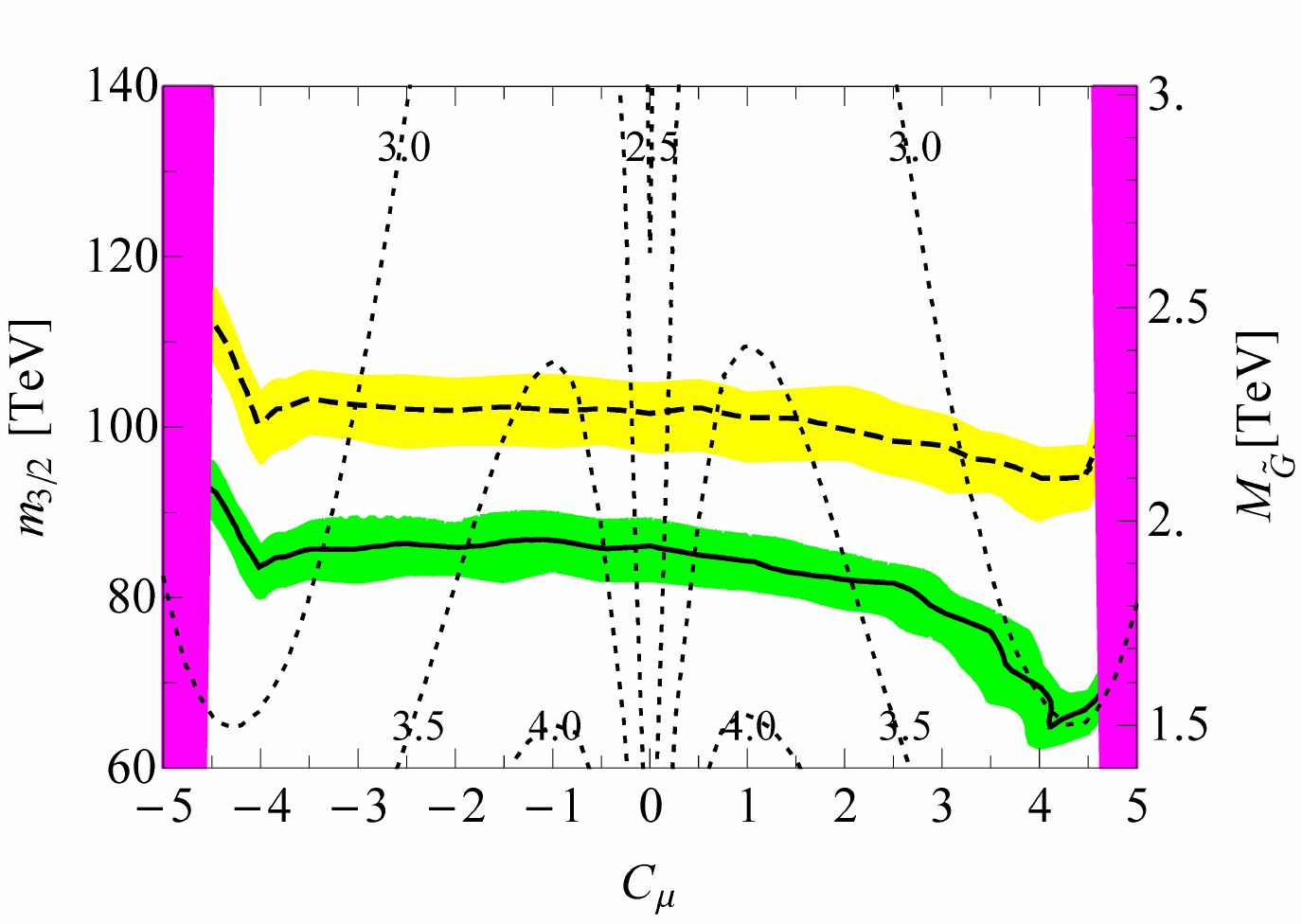}
                \label{Plot2d}
        \end{subfigure}
				\caption{5$\sigma$ discovery limits for anomaly mediation at LHC 14 for (solid) 300 $\text{fb}^{-1}$ and (dashed) 3000 $\text{fb}^{-1}$ integrated luminosity. The green band corresponds to the $1\sigma$ uncertainty on the gluino pair production cross-section for 300 $\text{fb}^{-1}$, the yellow band corresponds to the $1\sigma$ uncertainty on the gluino pair production cross-section for 3000 $\text{fb}^{-1}$, and the purple bands are the forbidden region of color-breaking vacuum. Contour lines of constant $M_{\tilde{B}}$, $M_{\tilde{W}}$, $\mu$, and $\tan\beta$ are shown respectively in (a), (b), (c), and (d).}\label{AnomalyLHC145s}
\end{figure}

\begin{figure}
        \centering
        \begin{subfigure}[b]{0.5\textwidth}
                \centering
                \caption{$M_{\tilde{B}}$ [TeV]}
                \includegraphics[width=\textwidth,bb=0 0 400 275]{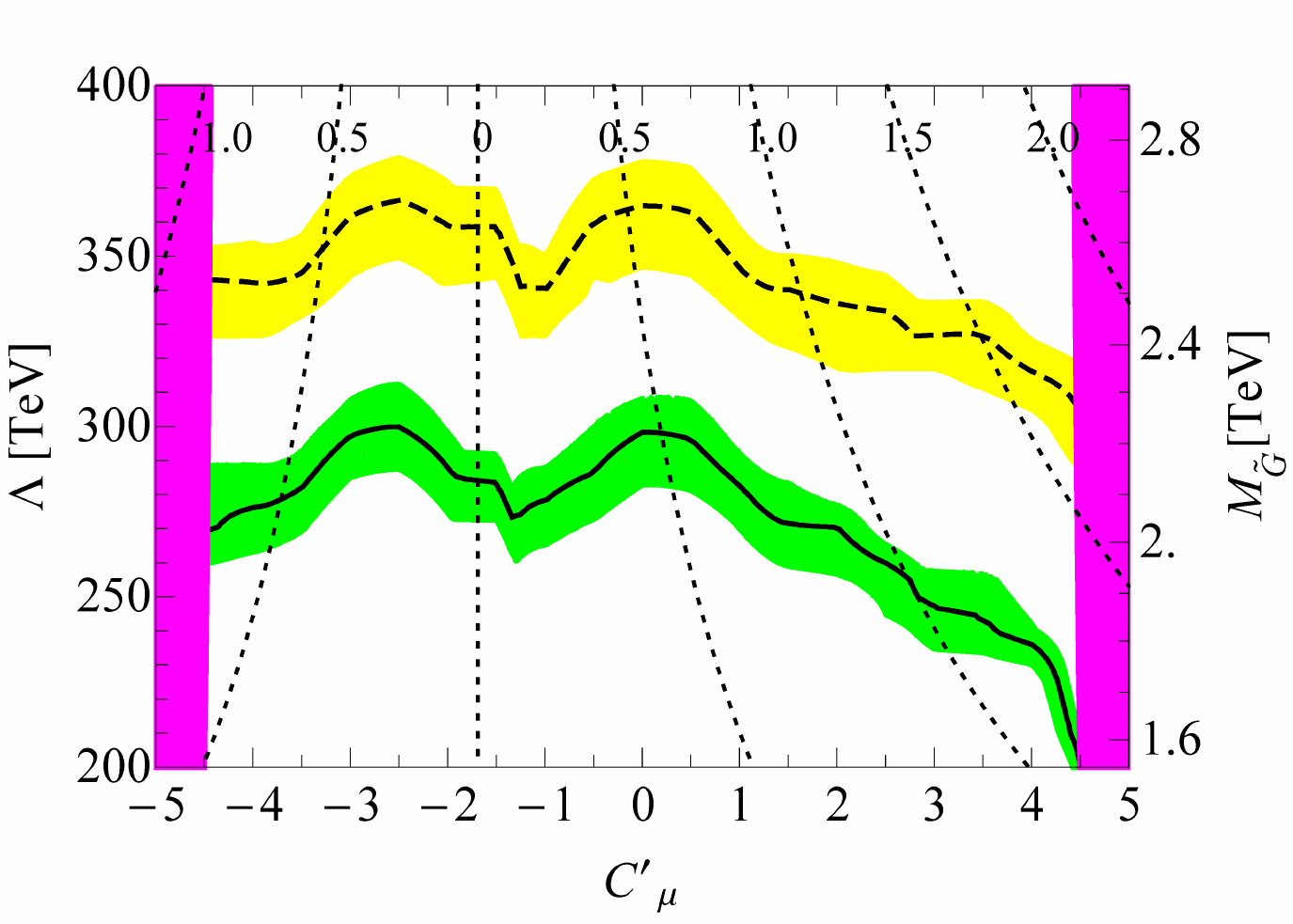}
                \label{Plot2a}
        \end{subfigure}%
        ~
        \begin{subfigure}[b]{0.5\textwidth}
                \centering
                \caption{$M_{\tilde{W}}$ [TeV]}      
                \includegraphics[width=\textwidth,bb=0 0 400 275]{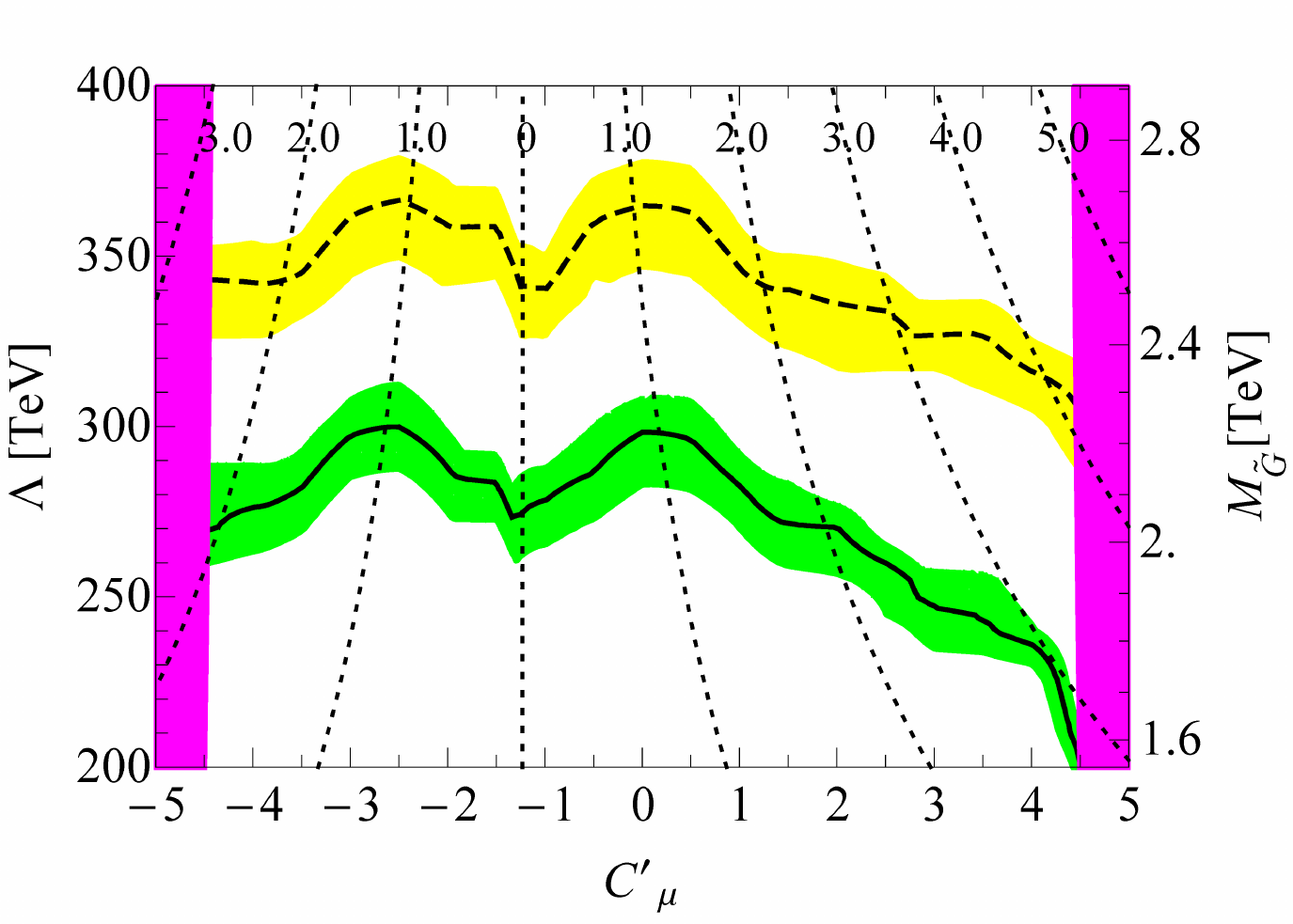}
                \label{Plot2b}
        \end{subfigure}
        \begin{subfigure}[b]{0.5\textwidth}
                \centering
                \caption{$\mu$ [PeV]}
                \includegraphics[width=\textwidth,bb=0 0 400 275]{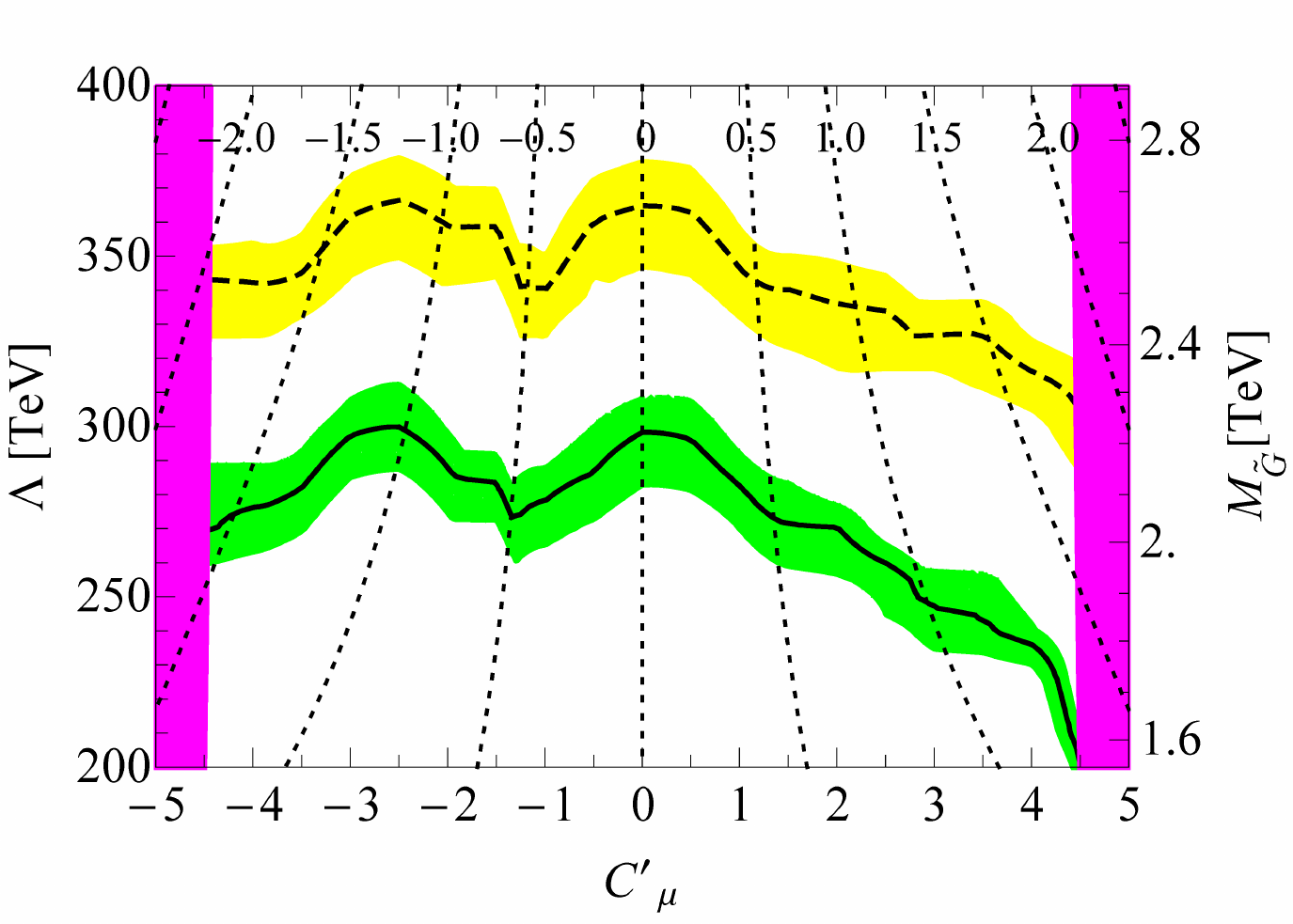}
                \label{Plot2c}
        \end{subfigure}%
        ~
        \begin{subfigure}[b]{0.5\textwidth}
                \centering
                \caption{$\tan\beta$}      
                \includegraphics[width=\textwidth,bb=0 0 400 275]{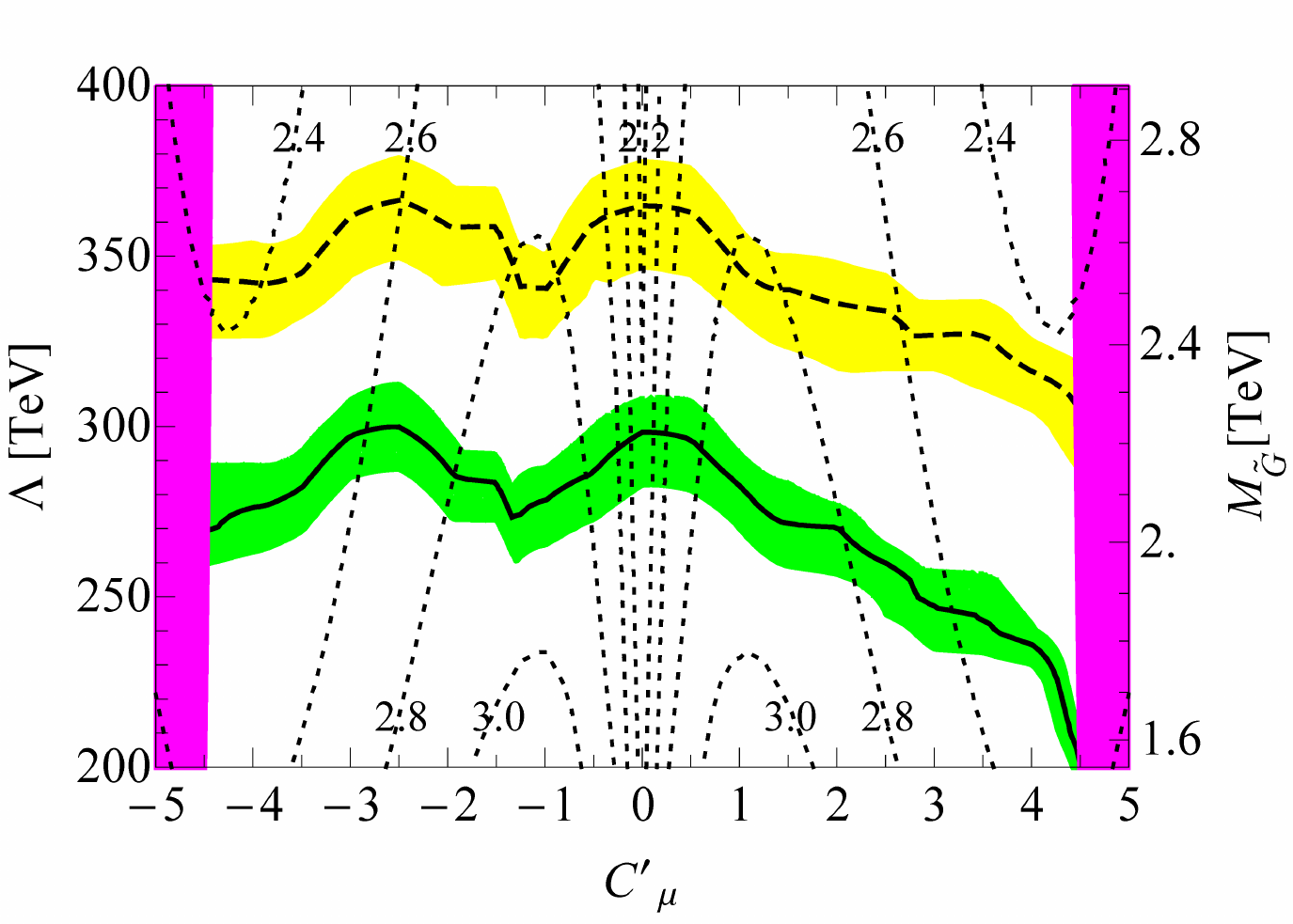}
                \label{Plot2d}
        \end{subfigure}
				\caption{95$\%$ exclusion limits for gauge mediation at LHC 14 for (solid) 300 $\text{fb}^{-1}$ and (dashed) 3000 $\text{fb}^{-1}$ integrated luminosity. The green band corresponds to the $1\sigma$ uncertainty on the gluino pair production cross-section for 300 $\text{fb}^{-1}$, the yellow band corresponds to the $1\sigma$ uncertainty on the gluino pair production cross-section for 3000 $\text{fb}^{-1}$, and the purple bands are the forbidden region of color-breaking vacuum. Contour lines of constant $M_{\tilde{B}}$, $M_{\tilde{W}}$, $\mu$, and $\tan\beta$ are shown respectively in (a), (b), (c), and (d).}\label{GaugeLHC1495}
\end{figure}

\begin{figure}
        \centering
        \begin{subfigure}[b]{0.5\textwidth}
                \centering
                \caption{$M_{\tilde{B}}$ [TeV]}
                \includegraphics[width=\textwidth,bb=0 0 400 275]{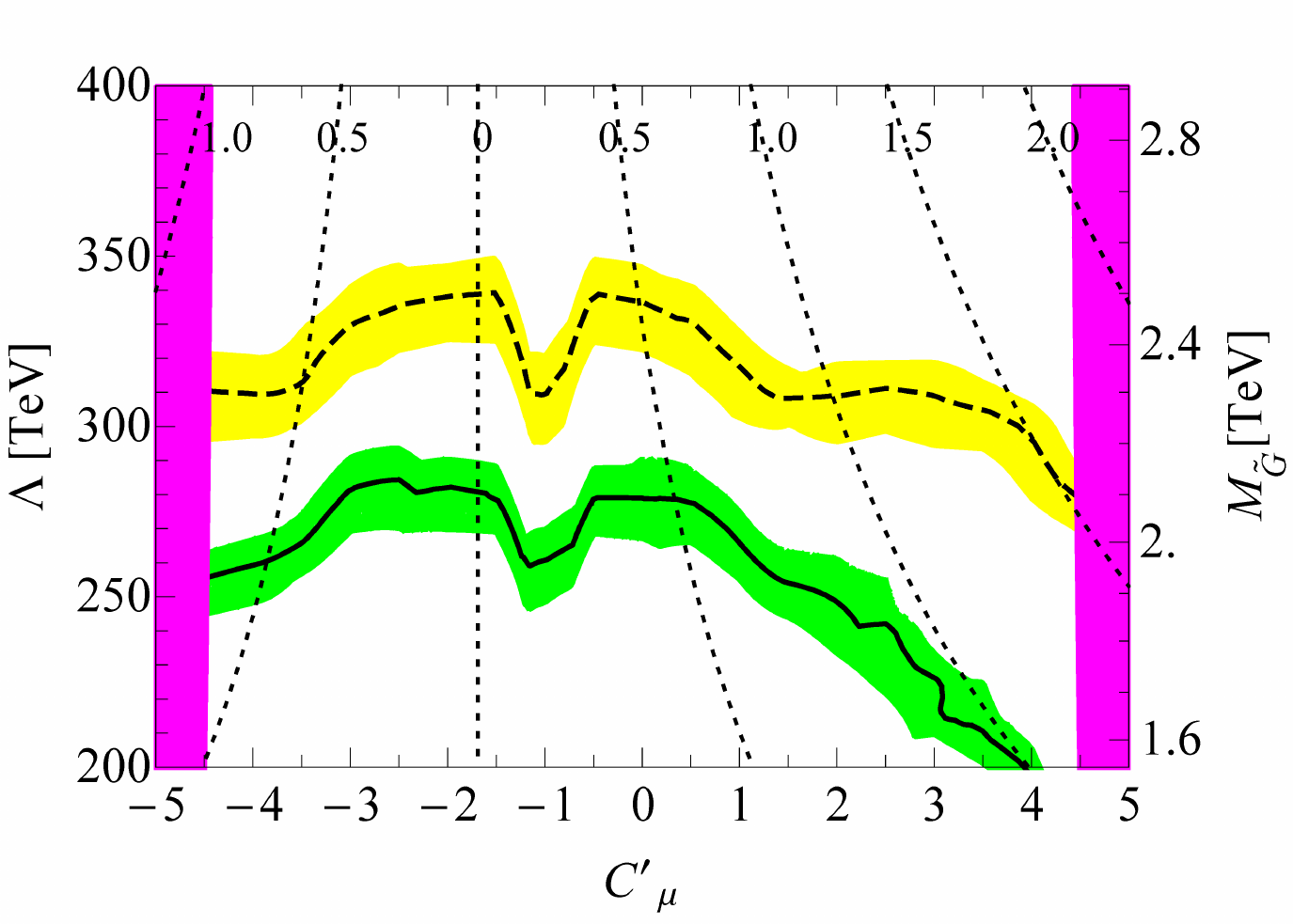}
                \label{Plot2a}
        \end{subfigure}%
        ~
        \begin{subfigure}[b]{0.5\textwidth}
                \centering
                \caption{$M_{\tilde{W}}$ [TeV]}      
                \includegraphics[width=\textwidth,bb=0 0 400 275]{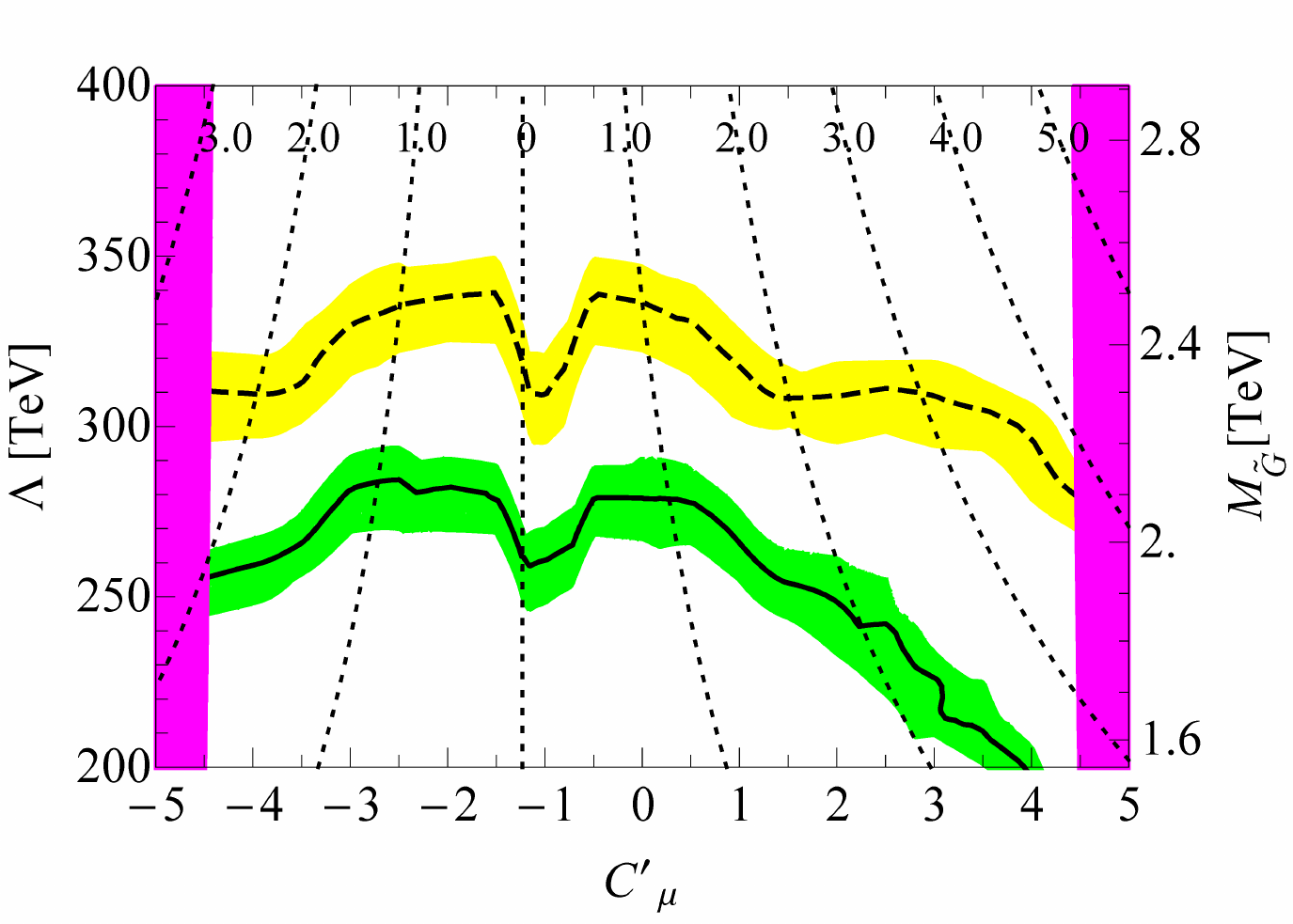}
                \label{Plot2b}
        \end{subfigure}
        \begin{subfigure}[b]{0.5\textwidth}
                \centering
                \caption{$\mu$ [PeV]}
                \includegraphics[width=\textwidth,bb=0 0 400 275]{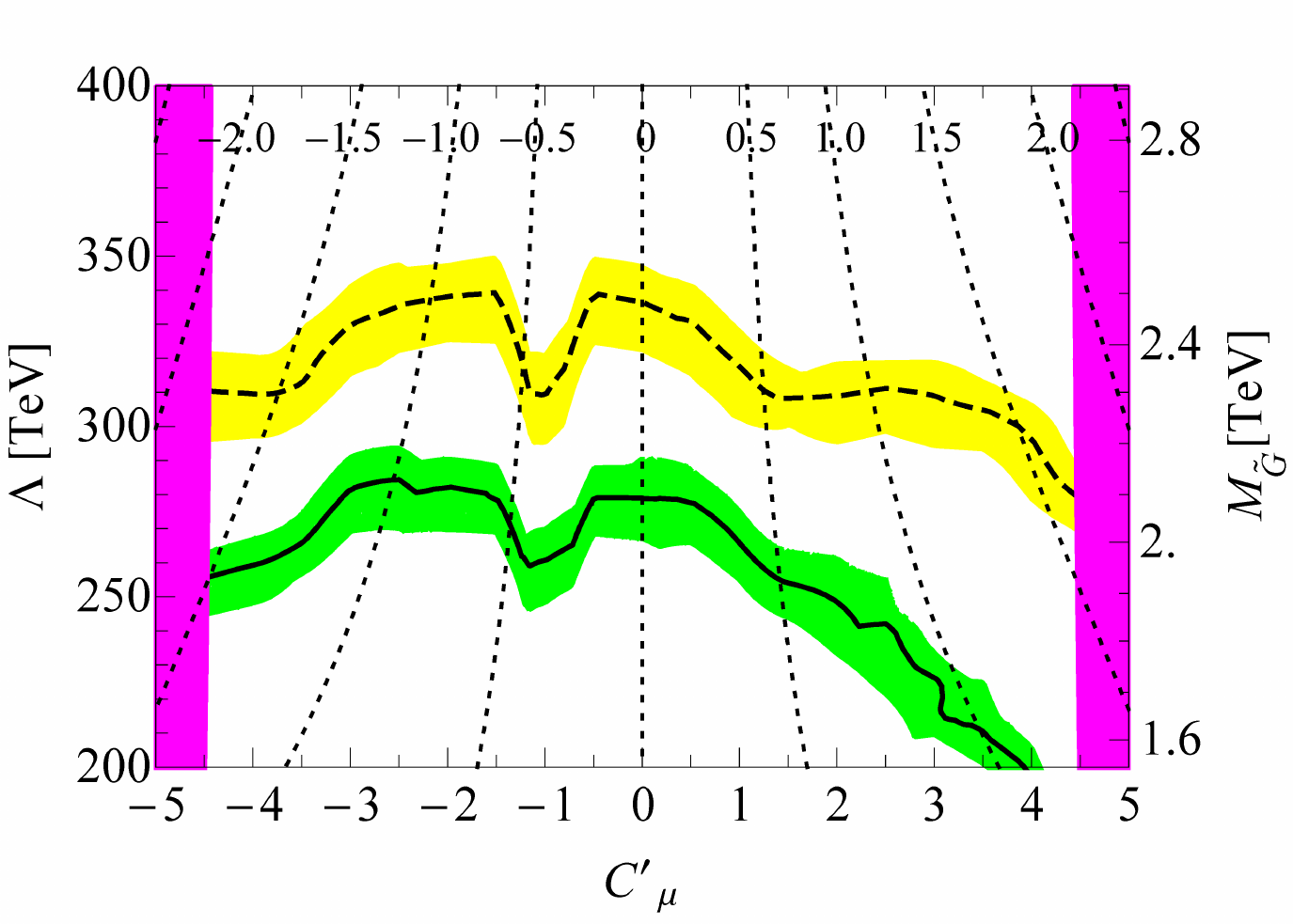}
                \label{Plot2c}
        \end{subfigure}%
        ~
        \begin{subfigure}[b]{0.5\textwidth}
                \centering
                \caption{$\tan\beta$}      
                \includegraphics[width=\textwidth,bb=0 0 400 275]{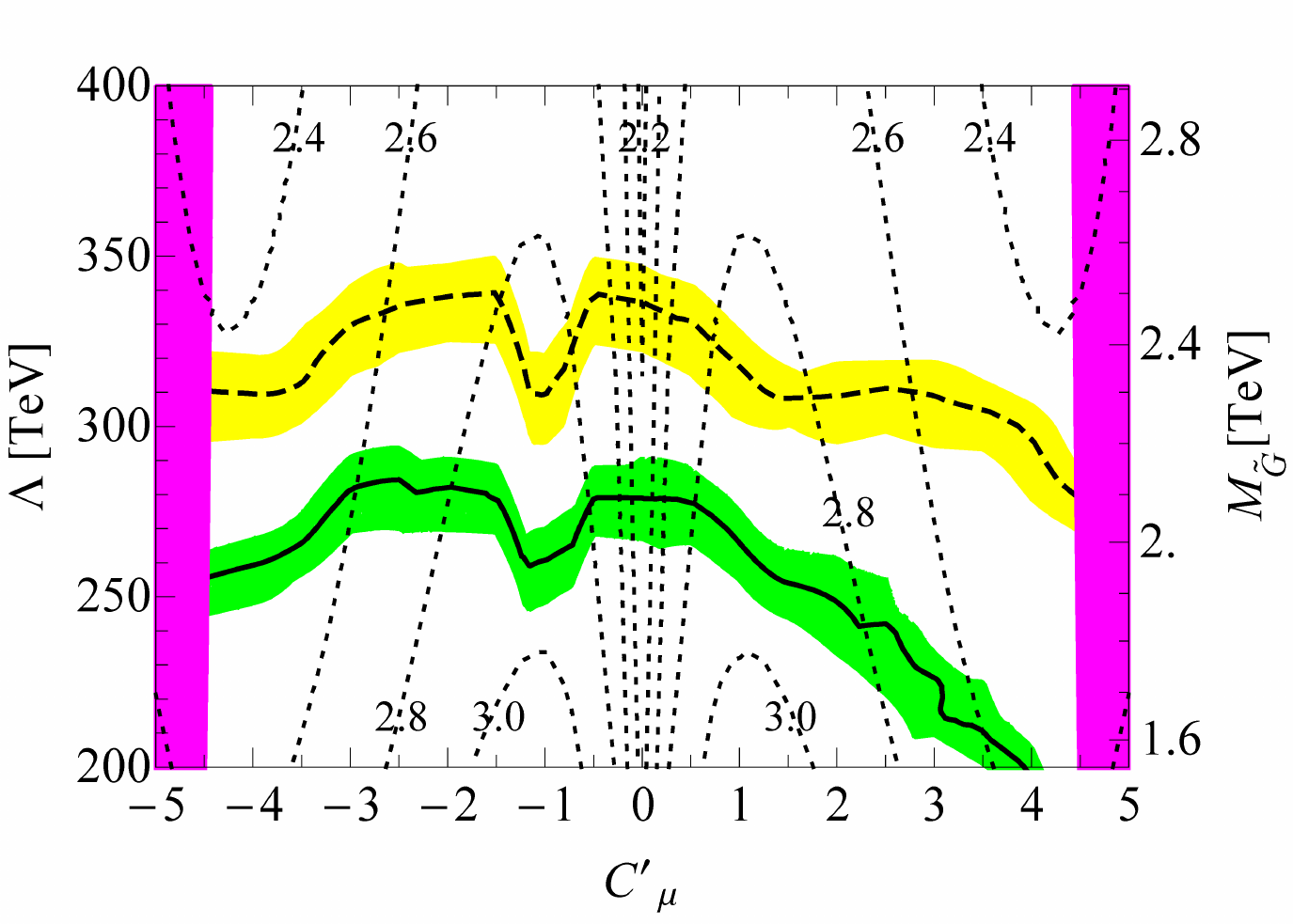}
                \label{Plot2d}
        \end{subfigure}
				\caption{5$\sigma$ discovery limits for gauge mediation at LHC 14 for (solid) 300 $\text{fb}^{-1}$ and (dashed) 3000 $\text{fb}^{-1}$ integrated luminosity. The green band corresponds to the $1\sigma$ uncertainty on the gluino pair production cross-section for 300 $\text{fb}^{-1}$, the yellow band corresponds to the $1\sigma$ uncertainty on the gluino pair production cross-section for 3000 $\text{fb}^{-1}$, and the purple bands are the forbidden region of color-breaking vacuum. Contour lines of constant $M_{\tilde{B}}$, $M_{\tilde{W}}$, $\mu$, and $\tan\beta$ are shown respectively in (a), (b), (c), and (d).}\label{GaugeLHC145s}
\end{figure}

The results can be seen for anomaly mediation in figures \ref{AnomalyLHC1495} and \ref{AnomalyLHC145s} for 95$\%$ exclusion and $5\sigma$ discovery respectively, as well as for gauge mediation in figure \ref{GaugeLHC1495} and \ref{GaugeLHC145s} for 95$\%$ exclusion and $5\sigma$ discovery respectively. The curves are essentially scaled up versions of the 8 TeV constraints. The anomaly mediation limits curves are flatter than those for the current LHC constraints. This can be explained by the fact that the branching ratio to the LSP and two tops decreases more slowly as $C_{\mu}$ increases because heavier gluinos are being probed.

\subsection{Prospects at a 100 TeV collider}\label{SectionProspects}
To fully explore the possibility of discovering Split supersymmetry at colliders, we study the prospect of a 100 TeV collider following the same procedure as in the previous two sections. Our high MET cuts are adapted from \cite{Jung:2013zya}, which are themselves based on \cite{Hinchliffe:2002mn}. These cuts rely on $M_{\text{eff}}$ which is defined as
\begin{equation}
   M_{\text{eff}}=\sum_i p_T(i)+\text{MET}.
\end{equation}
The sum is on jets with $p_T > 50$ GeV and $|\eta| < 5$ and leptons with $p_T > 15$ GeV and $|\eta| < 2.5$. We push things further than \cite{Jung:2013zya} by requiring b-jets, implementing detector simulations, and using a set of signal regions optimized for different regions of parameter space. The preselection cuts are given by \cite{Jung:2013zya}
\begin{itemize}
   \item{Lepton veto,}
   \item{At least two jets with $p_T > 0.1 M_{\text{eff}}$,}
   \item{$\text{MET} > 0.2 M_{\text{eff}}$,}
   \item{$p_T(j_1) < 0.35 M_{\text{eff}}$,}
   \item{$\Delta \phi(j_1,\text{MET}) < \pi -0.2$,}
   \item{$\Delta \phi(j_1,j_2) < 2\pi/3$. }
\end{itemize}
The different signal regions correspond to different combinations of minimum b-jets requirements and $M_{\text{eff}}$ cuts and are given in table \ref{TableSRleptonveto}. 

\begin{table}[t!]
  \begin{center}
	\begin{tabular}{ |C{3cm}|C{2cm}|C{2cm}|C{3cm}|}
		\hline
    SR			& b-jets 		& $M_{\text{eff}}$ [TeV]	& Background		\\ \hline
		hMETb3A	& $\ge$ 3		&				$>$ 15.0					&	23.4				  \\ \hline
		hMETb3B & $\ge$ 3		&				$>$ 17.5					&	7.8					  \\ \hline
		hMETb3C	& $\ge$ 3		&				$>$ 20.0					&	2.3  					\\ \hline
		hMETb4A	& $\ge$ 4		&				$>$ 12.5					&	12.6 					\\ \hline
		hMETb4B	& $\ge$ 4		&				$>$ 15.0					&	3.8 					\\ \hline
		hMETb4C	& $\ge$ 4		&				$>$ 17.5					&	1.5 					\\ \hline
		hMETb4D	& $\ge$ 4		&				$>$ 20.0					&	0.5						\\ \hline
	\end{tabular}%
	\caption{Signal regions for high MET. The background for 3 $\text{ab}^{-1}$ is also included.}\label{TableSRleptonveto}
	\end{center}
\end{table}

The SSDL cuts and the corresponding backgrounds are taken directly from \cite{Cohen:2013xda} and correspond to their search for gluino-neutralino model with heavy flavor decays. We verified that we could reproduce their results.

The detector card for Delphes is the standard 100 TeV card from Snowmass \cite{Anderson:2013kxz}. The background estimates for high MET are again obtained from the Snowmass online backgrounds \cite{Avetisyan:2013onh}. The backgrounds for the high MET signal regions are shown in table \ref{TableSRleptonveto} for  3 $\text{ab}^{-1}$ integrated luminosity. A 20$\%$ systematic uncertainty on all backgrounds is assumed \cite{Cohen:2013xda}. The discussion of pile-up for high MET or SSDL from the previous section still holds. We concentrate on the 0 pile-up case, as the average pile-up of a future 100 TeV is still unknown and as it only has a non-negligible effect on a small portion of our parameter space. The gluino pair production cross-section is calculated using NLL-fast \cite{Beenakker:1996ch, Kulesza:2008jb, Kulesza:2009kq, Beenakker:2009ha, Beenakker:2011fu} customized for a 100 TeV collider.

The results are again scaled up versions of LHC constraints with possible exclusion of up to a 14 TeV gluino in a large region of parameter space and discovery of up to 12 TeV. These numbers are similar to those obtained by \cite{Jung:2013zya}  which seem somewhat more optimistic (with a possible discovery of up to $\sim 15$ TeV).\footnote{ This might be due, for example, to the fact that we have used a detector simulation, but we haven't  directly checked that hypothesis.} For anomaly mediation, exclusion limits are governed by high MET signal regions and are thus very high until $C_\mu$ reaches 1. At this point, the spectrum becomes compact and the limits drop. The SSDL bins then dominate and the limits stabilize with a discovery reach of about $7$ TeV (this number is in fact quite close to the result of \cite{Cohen:2013xda}).  The exact same thing happens in the case of gauge mediation, except that the limits drop at $C'_\mu$ equal to 0.

\begin{figure}
        \centering
        \begin{subfigure}[b]{0.5\textwidth}
                \centering
                \caption{$M_{\tilde{B}}$ [TeV]}
                \includegraphics[width=\textwidth,bb=0 0 400 275]{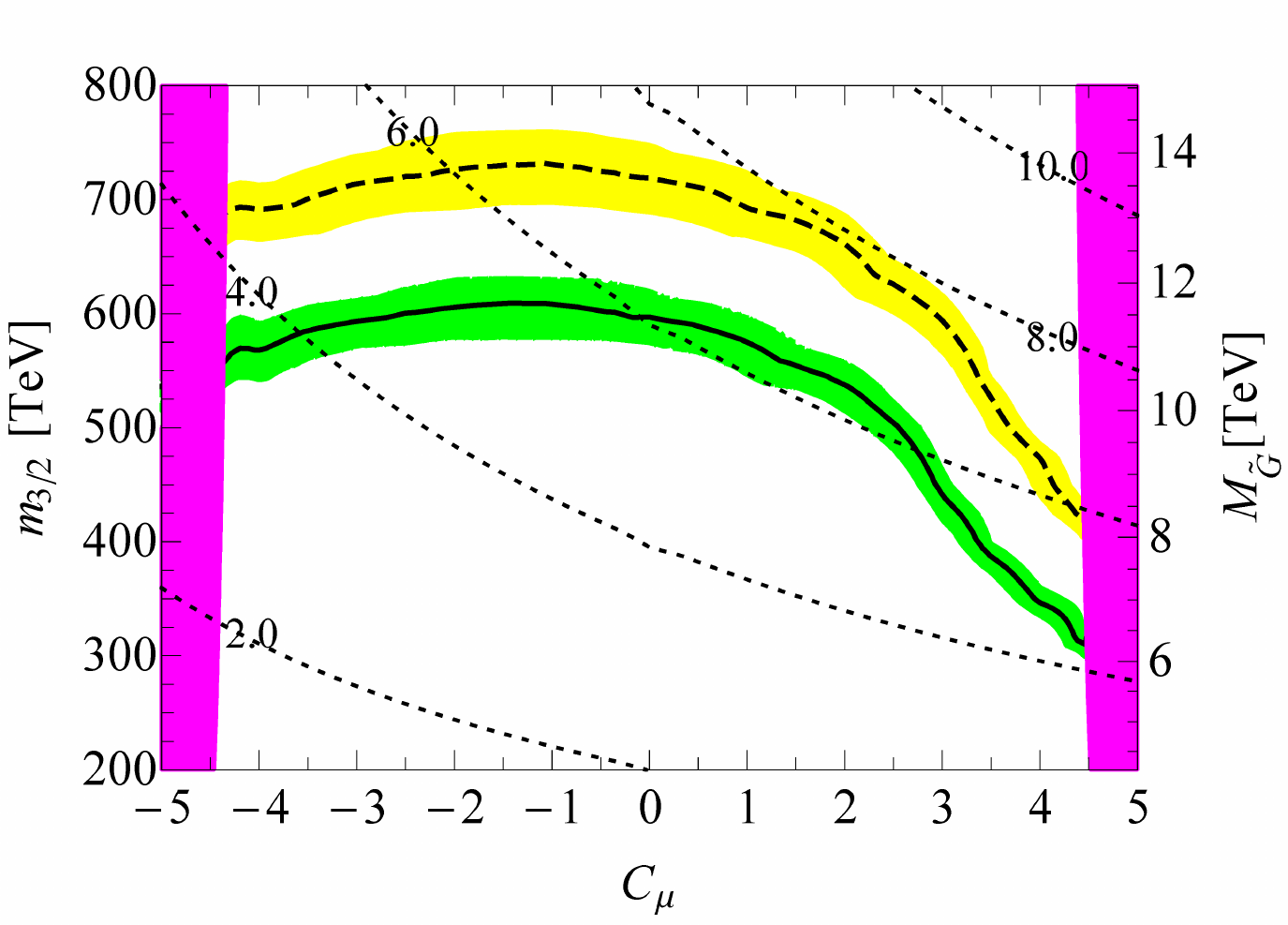}
                \label{Plot3a}
        \end{subfigure}%
        ~
        \begin{subfigure}[b]{0.5\textwidth}
                \centering
                \caption{$M_{\tilde{W}}$ [TeV]}      
                \includegraphics[width=\textwidth,bb=0 0 400 275]{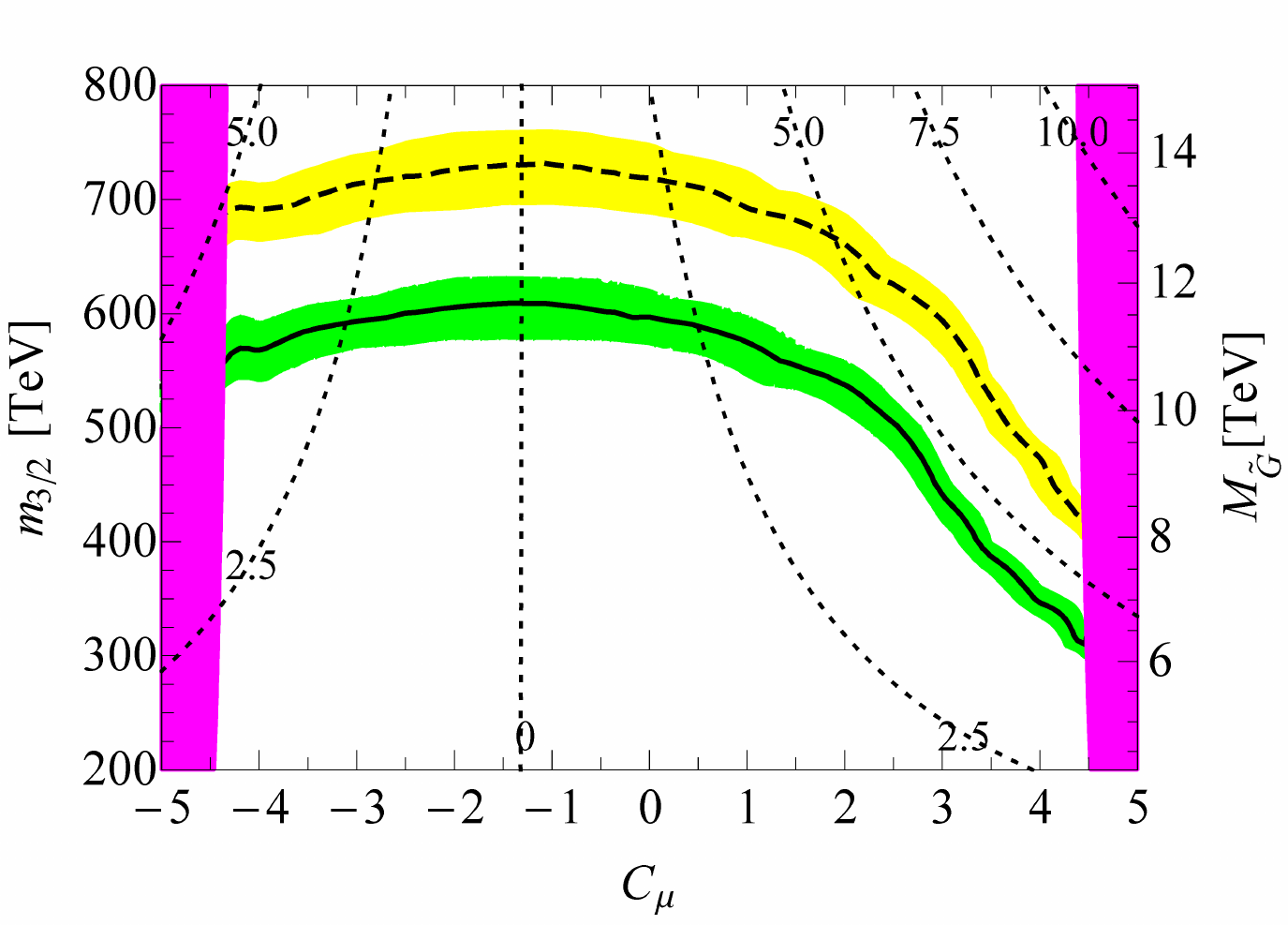}
                \label{Plot3b}
        \end{subfigure}
        \begin{subfigure}[b]{0.5\textwidth}
                \centering
                \caption{$\mu$ [PeV]}
                \includegraphics[width=\textwidth,bb=0 0 400 275]{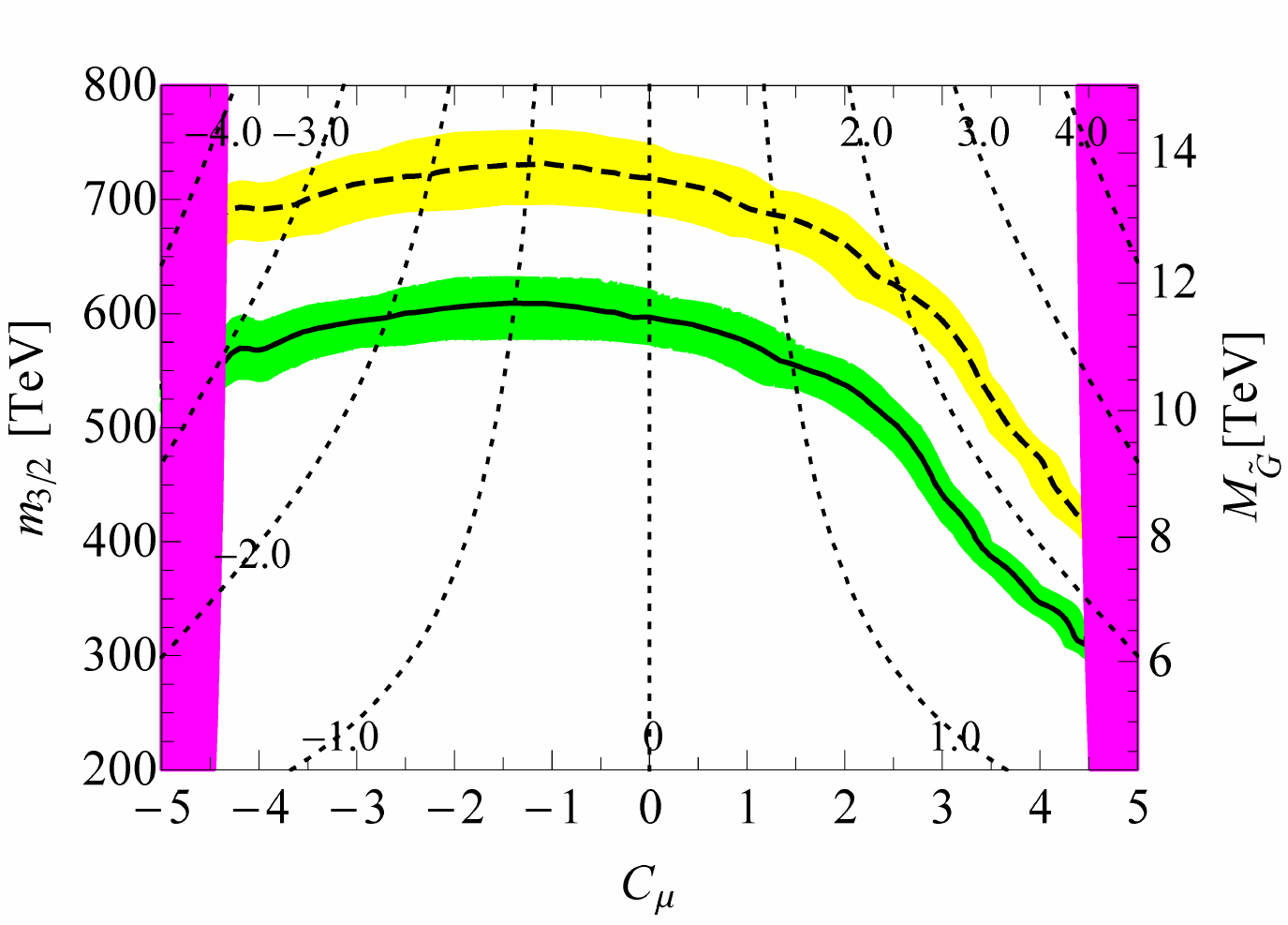}
                \label{Plot3c}
        \end{subfigure}%
        ~
        \begin{subfigure}[b]{0.5\textwidth}
                \centering
                \caption{$\tan\beta$}      
                \includegraphics[width=\textwidth,bb=0 0 400 275]{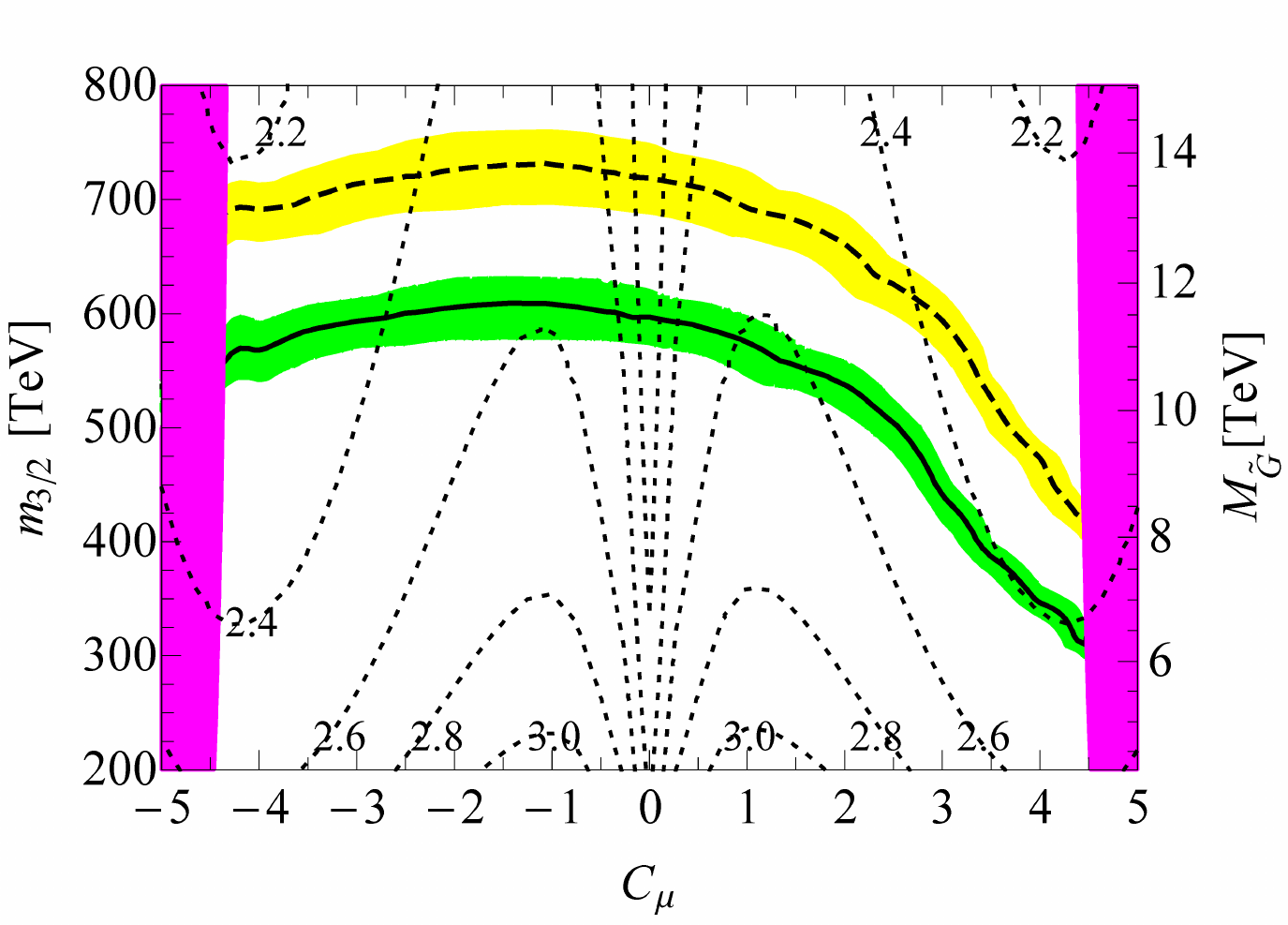}
                \label{Plot3d}
        \end{subfigure}
				\caption{95$\%$ (dashed) exclusion and $5\sigma$ (solid) discovery limits for anomaly mediation at a 100 TeV $pp$ collider with 3 $\text{ab}^{-1}$ integrated luminosity. The yellow band corresponds to the $1\sigma$ uncertainty on the gluino pair production cross-section for 95$\%$ exclusion, the green band corresponds to the $1\sigma$ uncertainty on the gluino pair production cross-section for $5\sigma$ discovery, and the purple bands are the forbidden region of color-breaking vacuum. Contour lines of constant $M_{\tilde{B}}$, $M_{\tilde{W}}$, $\mu$, and $\tan\beta$ are shown respectively in (a), (b), (c), and (d).}\label{Anomaly100}
\end{figure}

\begin{figure}
        \centering
        \begin{subfigure}[b]{0.5\textwidth}
                \centering
                \caption{$M_{\tilde{B}}$ [TeV]}
                \includegraphics[width=\textwidth,bb=0 0 400 275]{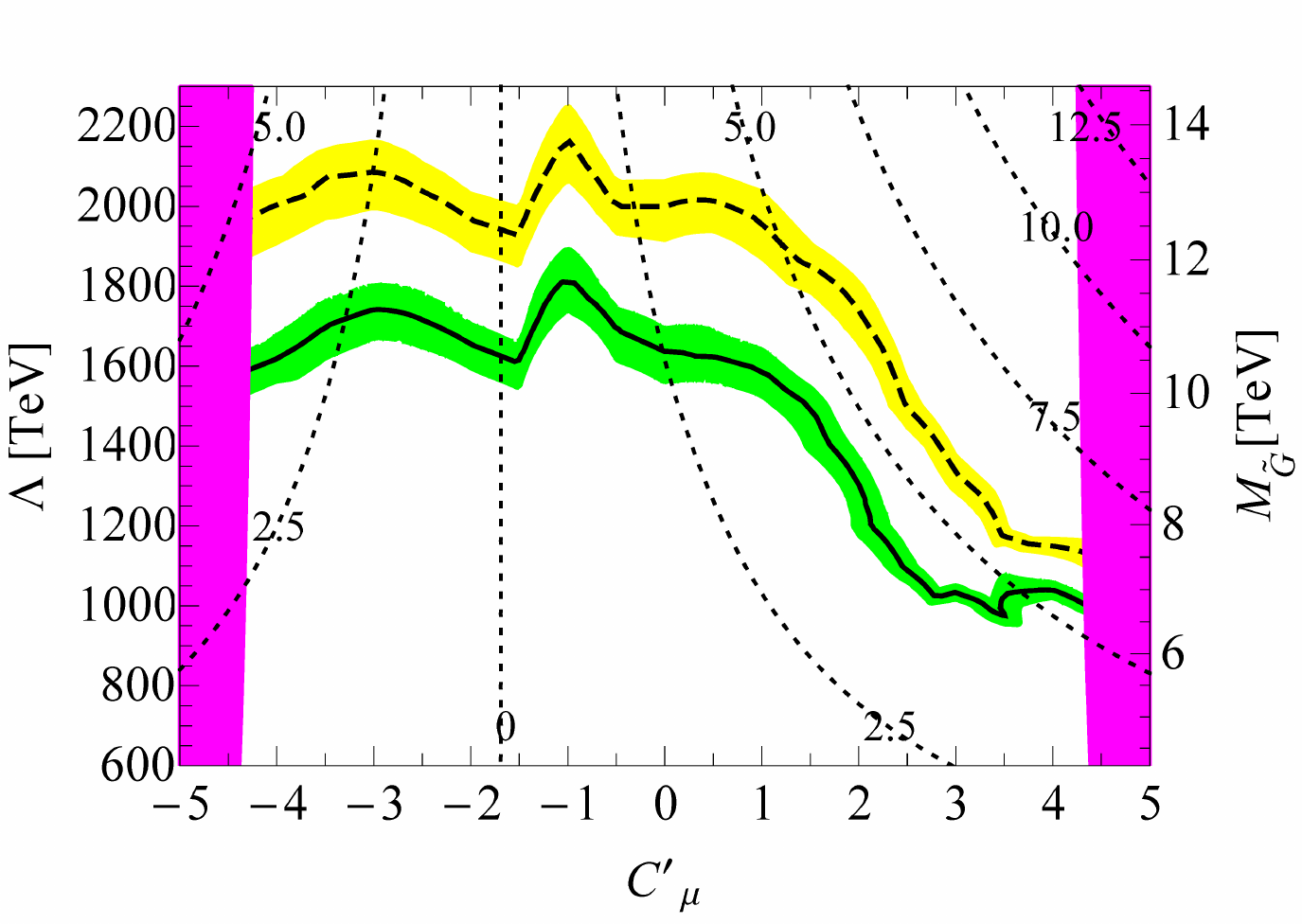}
                \label{Plot4a}
        \end{subfigure}%
        ~
        \begin{subfigure}[b]{0.5\textwidth}
                \centering
                \caption{$M_{\tilde{W}}$ [TeV]}      
                \includegraphics[width=\textwidth,bb=0 0 400 275]{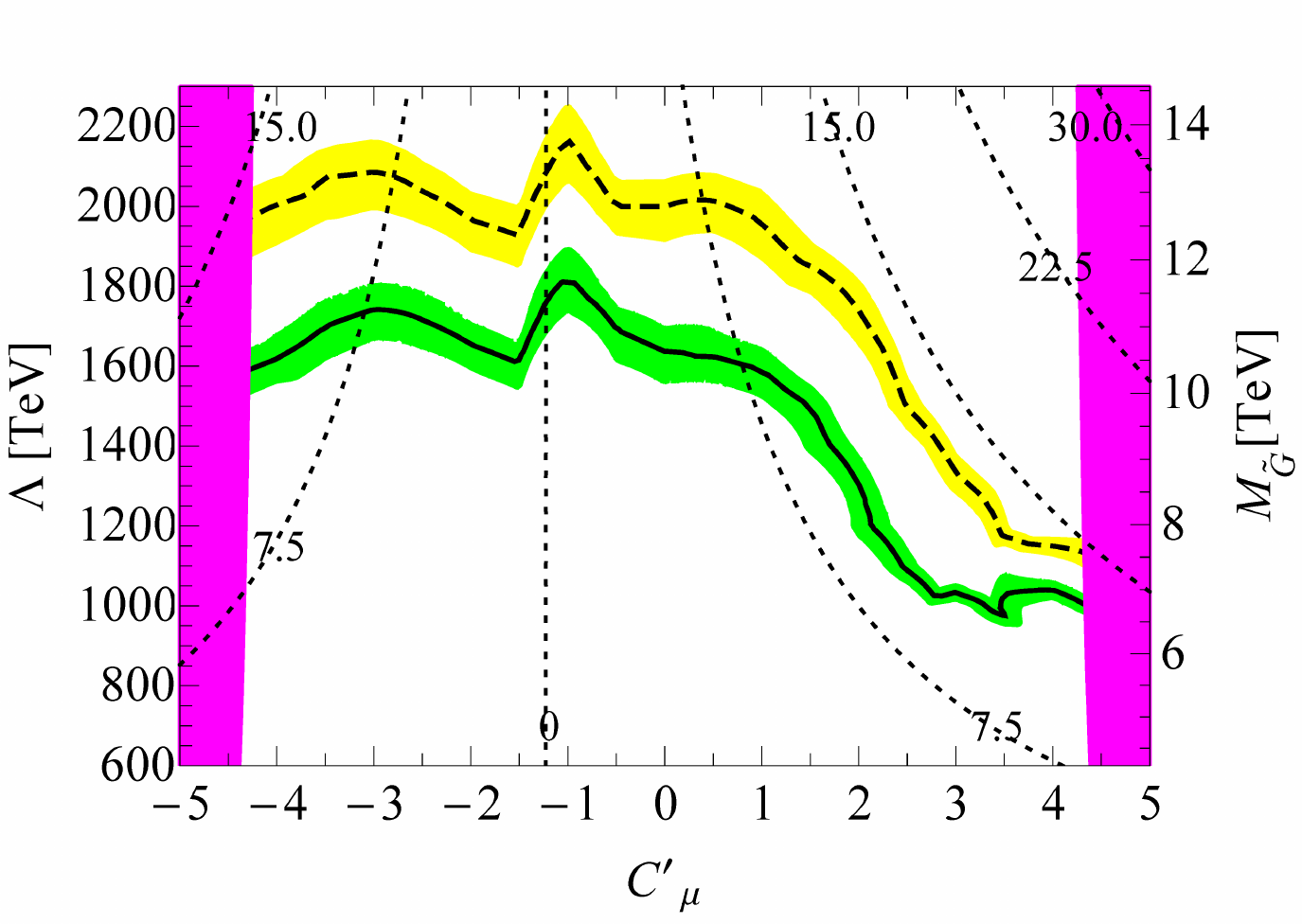}
                \label{Plot4b}
        \end{subfigure}
        \begin{subfigure}[b]{0.5\textwidth}
                \centering
                \caption{$\mu$ [PeV]}
                \includegraphics[width=\textwidth,bb=0 0 400 275]{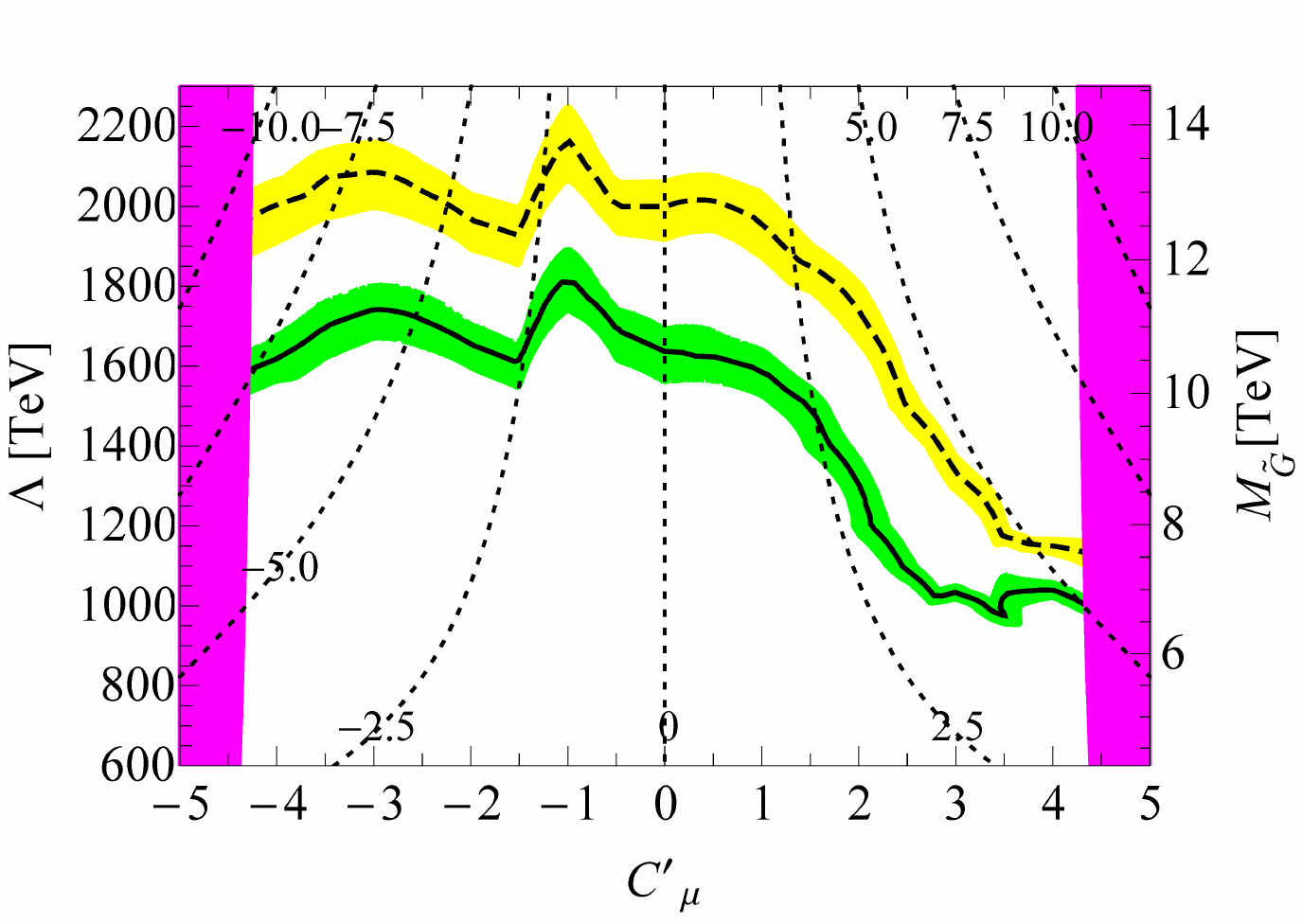}
                \label{Plot4c}
        \end{subfigure}%
        ~
        \begin{subfigure}[b]{0.5\textwidth}
                \centering
                \caption{$\tan\beta$}      
                \includegraphics[width=\textwidth,bb=0 0 400 275]{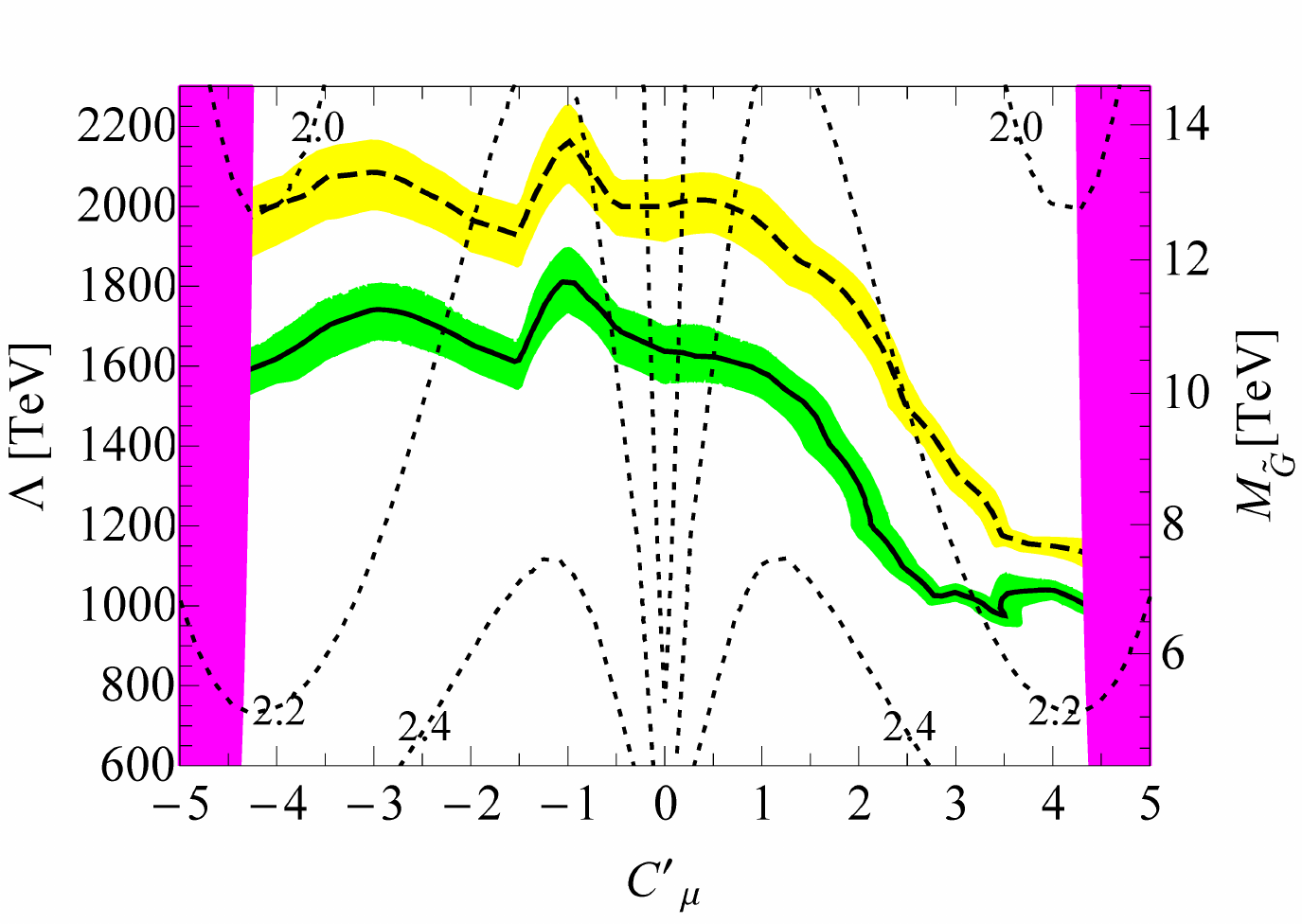}
                \label{Plot4d}
        \end{subfigure}
				\caption{95$\%$ (dashed) exclusion and $5\sigma$ (solid) discovery limits for gauge mediation at a 100 TeV $pp$ collider with 3 $\text{ab}^{-1}$ integrated luminosity. The yellow band corresponds to the $1\sigma$ uncertainty on the gluino pair production cross-section for 95$\%$ exclusion, the green band corresponds to the $1\sigma$ uncertainty on the gluino pair production cross-section for $5\sigma$ discovery, and the purple bands are the forbidden region of color-breaking vacuum. Contour lines of constant $M_{\tilde{B}}$, $M_{\tilde{W}}$, $\mu$, and $\tan\beta$ are shown respectively in (a), (b), (c), and (d).}\label{Gauge100}
\end{figure}

\section{Conclusions}
In light of ever stronger constraints from collider physics, Mini-Split scenarios become more and more appealing. In these models, a small hierarchy exists between the sfermions and gauginos, with the gauginos being near the electroweak scale. This kind of spectrum could easily arise from anomaly mediation and also from gauge mediation. In these models the electroweak scale is tuned, but gauge couplings could still unify at a high scale, and the models have possible dark matter candidates. The hierarchy between the scalars and the gauginos leads to large radiative corrections which can greatly modify the standard mass spectra of anomaly and gauge mediation.

In this paper we studied hadron collider constraints and prospects on these deflected anomaly mediation and deflected gauge mediation models. By using a simple parametrization of the models and assuming a lighter third generation and a heavy Higgsino, we recast SUSY searches from ATLAS and CMS to obtain exclusions on the parameter space of the models. The known mass of the Higgs boson and the absence of color-vacuum were also taken into account. Results for anomaly and gauge mediation can be seen respectively in figures \ref{AnomalyLHC} and \ref{GaugeLHC}. We also obtained future prospects for deflected anomaly mediation and deflected gauge mediation for LHC 14 and a 100 TeV collider. For LHC 14, the 95$\%$ projected exclusion limits are shown in figure \ref{AnomalyLHC1495} and \ref{GaugeLHC1495} for anomaly and gauge mediation respectively and the 5$\sigma$ discovery prospects are shown in \ref{AnomalyLHC145s} and \ref{GaugeLHC145s}. The prospects at a 100 TeV collider for anomaly and gauge mediation are found in figures \ref{Anomaly100} and \ref{Gauge100} respectively.

While the goal of this work was to explore the collider phenomenology of Mini-Split models, dark matter properties could also be used to further restrict the parameter space. The thermal abundance of the dark matter candidate is strongly dependent on the identity of the LSP. For a Wino LSP, the correct thermal relic abundance can be obtained for a wino mass of 2.7 TeV \cite{Bagnaschi:2014rsa,ArkaniHamed:2012gw}. This region of parameter space is not constrained by the LHC, but is within reach of a 100 TeV collider. Wino LSP with lighter mass could be accommodated by invoking non thermal production \cite{ArkaniHamed:2012gw}. Similarly, Bino LSP, which tend to overclose the universe, could be accommodated if there was late entropy production or a low reheating temperature.

\acknowledgments
This work was supported in part by the Natural Sciences and Engineering Research Council of Canada (NSERC). HB acknowledges support from the Ontario Graduate Scholarship (OGS).

\bibliographystyle{JHEP}
\bibliography{Paper1}

\end{document}